\documentclass[reqno,12pt]{article}
\usepackage{a4wide}
\usepackage{amssymb}
\usepackage{enumerate}{}
\usepackage{amsthm,amsfonts,amsmath}
\newcommand{\e}{\varepsilon}
\newcommand{\zg}{ z_\gamma}
\newcommand{\zga}{ z_{\gamma-\frac {1}{2}}}
\newcommand{\zgp}{ z'_\gamma}
\newcommand{\zgap}{ z'_{\gamma-\frac {1}{2}}}
\newcommand{\zhal}{(1+z^2)^{1/2}}
\newcommand{\zphal}{(1+z'^2)^{1/2}}
\newtheorem{theorem}{\bf Theorem}[section]
\newtheorem{lm}[theorem]{\bf Lemma}
\newcommand{\R}{{\mathord{\mathbb R}}}

\numberwithin{equation}{section}
\numberwithin{theorem}{section}

\let\e=\varepsilon

\let\g=\gamma

\let\r=\rho

\let\G=\Gamma

\let\s=\sigma
\newcommand{\1}{\,\rlap{\small 1}\kern.13em 1}
\newcommand{\la}{\langle}
\newcommand{\ra}{\rangle}
\newcommand{\sqr}[2]{{\vcenter{\hrule height.#2pt%
                      \hbox{\vrule width.#2pt height#1pt\kern#1pt%
                            \vrule width.#2pt}%
                      \hrule height.#2pt}}}

\def\O{\Omega}

\begin{document}

\title{\bf{Stability of the Front under a Vlasov-Fokker-Planck Dynamics}}

\author{R. Esposito$^1$, Y. Guo$^2$ and R. Marra$^3$\\  \footnotesize{$^1$Dipartimento di
Matematica, Universit\`a di L'Aquila, Coppito, 67100 AQ, Italy}
\\ \footnotesize{$^2$Division of Applied Mathematics,
Brown University, Providence, RI 02812,  U.S.A. }
\\ \footnotesize{$^3$Dipartimento di Fisica and Unit\`a INFN, Universit\`a di Roma Tor Vergata, 00133 Roma, Italy.}
}

\maketitle

\begin{abstract}

We consider a kinetic model for a system of two species of particles interacting through a long
range repulsive potential and a reservoir at given temperature. The model is described by a set of two coupled Vlasov-Fokker-Plank equations. The important front solution, which represents the phase  boundary, is a one-dimensional stationary solution on the real line with given asymptotic values at infinity. We prove the asymptotic stability of the front for small symmetric perturbations.
\end{abstract}

\bigskip

Keywords: {\it Vlasov, Fokker-Planck, phase transition, stability, fronts}

\section{Introduction and Notations.}

The dynamical study of phase transitions has been tackled, among the others, with an approach based on kinetic equations modeling short range and long range interactions which are responsible of critical behaviors. An example of such models has been proposed in \cite{BL} where the authors study a system
of two species of particles undergoing collisions regardless of the species and interacting via long range repulsive forces between different species. A simplification of such model has been considered  in \cite{MM} where a kinetic model has been introduced for a system of two species of particles interacting through a long
range repulsive potential and with a reservoir at a given temperature $T$.  The interaction with the reservoir is modeled by a Fokker-Plank operator and the interaction between the two species by
 a Vlasov force. The system is described by the one-particle distribution functions $f_i(x,v,t)$, $i=1,2$, with  $(x,v)\in\Omega\times \mathbb{R}^3$ the position and velocity of the particles. The distribution functions $f_i$ are solutions of a system of two coupled Vlasov-Fokker-Plank (VFP) equations in a domain $\Omega\subset \mathbb{R}^3$:

\begin{equation}
\displaystyle{\partial_{t}f_i+v\cdot\nabla_xf_i+F_i\cdot\nabla_vf_i=Lf_i,}
\label{VFP}
\end{equation}
where
\begin{equation}
Lf_i=\nabla_v\cdot\left(M\nabla_v\left(\frac{f_i}{M}\right)\right),
\label{fp}
\end{equation}
and $M$ is the  Maxwellian 
\[
M=\left(\frac{\beta}{2\pi}\right)^{\frac{3}{2}}e^{-\frac{\beta}{2}v^2}
\]
with mean zero and variance $\beta^{-1}=T$ which is interpreted as the temperature of the thermal reservoir. The self-consistent Vlasov
force, representing the repulsion between particles of different species, is
\[
F_i=-\nabla_x\int_{\Omega} {\rm d }x'U(|x-x'|)\int_{\mathbb R^3} {\rm d} vf_{j}(x',v, t)\ ,\]
with $j=i+1\pmod 2$ (this notation will be used in the rest of the paper). The potential function $U$ is a positive, bounded, smooth, monotone decreasing function on $\mathbb{R}_+$ , with compact support and $\int_{\mathbb{R}^3} {\rm d}x U(|x|)=1$. There is a natural Liapunov functional, the {\it free energy}, for this dynamics,
\begin{eqnarray*}
\mathcal G (f_1,f_2) :&=& \int _{{\Omega}\times \mathbb R^3}{\rm d }x {\rm d }v\left [(f_1\ln f_1) + (f_2\ln
f_2)+\frac{\beta}{2}(f_1+f_2) v^2\right] \\ &+&{\beta\int _{{\Omega}\times \Omega}{\rm d}x{\rm
d}y U(|x-y|)\int_{\mathbb  R^3}{\rm d }v f_1(x,v)\int_{\mathbb  R^3} {\rm d
}v' f_2(y,v')}\ .\\ \label{freeg}
\end{eqnarray*}
In fact, we have that
\[\frac{d}{dt}\mathcal G (f_1,f_2)=-\sum_{i=1,2}\int _{{\Omega}\times \mathbb  R^3}{\rm d}x{\rm d}v\frac
{{M}^2}{f_i} \left|\nabla_v\frac{f_i}{M}\right|^2\le 0
\]
and the time derivative is zero if and only if $f_i$ are of the form $f_i=\rho_i M$, where $\rho_i$ are functions only of the position.   If we put these expressions back in the VFP equations we see that the stationary solutions of (\ref{VFP}) have densities satisfying the equations
\begin{equation}\label{statio}
\ln\rho_i(x)+\beta\int_{\Omega}{\rm d }x' U(|x-x'|)\rho_j(x')=C_i\; x\in\Omega,\; i=1,2
\end{equation}
and $C_i$ are arbitrary constants, related to the total masses $n_i|\Omega|$ of the components of the mixture. Moreover, replacing $f_i$ by $\rho_i M$ in the
functional $\mathcal G$ and integrating out the velocity variable we obtain a functional on the densities  $\rho_i $
\begin{equation}\label{free}
\mathcal{F}(\rho_1,\rho_2)=\int_{\Omega} {\rm d }x(\rho_1\ln\rho_1+\rho_2\ln\rho_2)+\beta\int _{\Omega\times \Omega}{\rm d }x {\rm d }yU(|x-y|)\rho_1(x)\rho_2(y)
\end{equation}
The Euler-Lagrange equations for the minimization of $\mathcal{F}$ with the constraint on the total masses 
\begin{equation}\label{constr}
\int_{\Omega}{\rm d }x \rho_i(x)= n_i  |\Omega|, \quad 1=1,2\ ,
\end{equation}
are exactly (\ref{statio}). We set $n=n_1+n_2$ the total average density.

In \cite{CCELM1} it is proved that for $n \beta\le 2$,  equations (\ref{statio}) in a torus have a unique homogeneous solution, while for $n \beta>2$  there are non homogeneous solutions. To explain the physical meaning of these non homogeneous solutions, we write the functional $\mathcal{F}(\rho_1,\rho_2)$ in the following equivalent form
\[\mathcal{F}(\rho_1,\rho_2)=\hskip -5.7pt\int_{\Omega} {\rm d }x f(\rho_1,\rho_2)+\frac{\beta}{2}\int_{\Omega^2} {\rm d }x {\rm d }y U(|x-y|)[\rho_1(x)-\rho_1(y)][\rho_2(y)-\rho_2(x)]\]
where $f(\rho_1,\rho_2)$  is the thermodynamic free energy  made of the entropy and the internal energy:
\[f(\rho_1,\rho_2)=\rho_1\log \rho_1+ \rho_2\log \rho_2+\beta \rho_1\rho_2\]
 The function $f(\rho_1,\rho_2)$ is not convex and has,  for any given temperature $T=\beta^{-1}$, two symmetric (under the exchange $1\to 2$) minimizers  if the total density $\displaystyle{n=\frac{1}{|\Omega|}\int_\O{\rm d }x(\rho_1+\rho_2)}$ is larger than a critical value $2T^{-1}$.
Indeed, there are two positive numbers $\rho^+>\rho^->0$ such that one minimizer is given by $\rho_1=\rho^+$, $\rho_2=\rho^-$ and the other by exchanging  the indices $1$ and $2$.  In other words, this system undergoes a first order phase transition with coexistence of two phases, one richer in the presence of species $1$ and the other richer in the presence of species $2$. 

If we look for the minimizers of the free energy functional $\mathcal{F}$ under the constraints (\ref{constr})
the only minimizers are homogeneous if we fix $(n_1,  n_2)$ equal to one of the two  minimizers of $f(\r_1,\r_2)$.

Otherwise, if $n_i\in (\rho^-,\rho^+)$,  $i=1,2$, below the critical value non homogeneous profiles may have lower free energy.
The structure of the minimizing profiles of density will be as close as possible to  one of the two minimizers of $f$: they will be close to one of the minimizing values in a region $B$, close to the other minimizing value in the complement but for a separating  region called {\it interface} where the minimizing profiles will interpolate  smoothly between the two values. A precise statement of this is proved in \cite{CCELM1} under the assumption that the size of $\Omega$ is large compared to the range of the potential $U$.

We can conclude then that the minimizers of $\mathcal G$ in a torus will be Maxwellians times densities $\rho_i$ of the form discussed above.
 Since $\mathcal G$  is a Liapunov functional, we expect that the minimizers are related to the stable solutions of the equations. In this paper we want to study the stability of the non homogeneous stationary solutions of the equations (\ref{VFP}), which are minimizers of the kinetic free energy $\mathcal{G}$.
 
Since planar interfaces play an important role in the study of the evolution of general interfaces, in this paper we focus on the so called {\it front} solutions, i.e. one-dimensional infinite volume solutions, with
\[x=(0,0,z)\qquad  -\infty<z<\infty\ .\]

The reason for choosing this setup is that in such a situation we know many more properties of the minimizers.  

To be more precise, we introduce the
 {\it excess free energy functional} in one dimension  on the infinite line defined as
\begin{equation}
\hat{\mathcal F}(\rho_1,\rho_2):=\lim_{N\to \infty}[\mathcal {F}_N(\rho_1,\rho_2)-\mathcal {F}_N(\rho^+, \rho^-)]
\label{excess}
\end{equation}
where $\mathcal {F}_N$ is the free energy associated to the interval $[-N,N]$ and $(\rho^+, \rho^-)$ is a homogeneous minimizer of the thermodynamic free energy $f$. We note that $\mathcal {F}_N(\rho^+, \rho^-)=\mathcal {F}_N(\rho^-, \rho^+)$. We look for the minimizers of the excess free energy such that
$\lim_{z\to\pm\infty}\rho_1(z)=\rho^\pm$, $\lim_{z\to\pm\infty}\rho_2(z)=\rho^\mp$, because otherwise the limit defining $\hat{\mathcal F}$ would not be finite. By the translation invariance of $\hat{\mathcal F}$ the minimizers are degenerate. We remove the degeneration by imposing the {\it centering condition}, $\rho_1(0)=\rho_2(0)$. In \cite{CCELM2}  it is proved that

\begin{theorem}\label{inf}
There exists a unique $C^\infty$ positive minimizer {\rm (front)} $w=(w_1 (z),w_2(z))$, with $w_1(z)=w_2(-z)$, for the one-dimensional excess free energy $\hat{\mathcal F}$, defined in (\ref{excess}),  in the class of continuous functions $\rho=(\rho_1,\rho_2)$ such that
\[\lim_{z\to\pm\infty}\rho_1= \rho^\pm,\quad \lim_{z\to\pm\infty}\rho_2= \rho^\mp.\]
The properties of the minimizer are: $w_1$ is monotone increasing and $w_2$ is monotone decreasing and
$$\rho^-<w_i(z)<\rho^+$$
for any $z\in \mathbb{R}$. 

Moreover, the front $w$ is smooth and satisfies the Euler-Lagrange equations (\ref{statio});
 its derivatives $w'$ satisfy the equations
\begin{equation}\label{e-l'}\frac{w_1'(z)}{w_1(z)}+\beta(U*w_2')(z)=0, \quad \frac{w_2'(z)}{w_2(z)}+\beta(U*w_1')(z)=0
\end{equation}
The front $w$ converges to its asymptotic values exponentially fast, in the sense that there is $\alpha>0$ such that
$$|w_1(z)-\rho_\mp|e^{\alpha|z|}\to 0 \text{ as } z\to \mp \infty, \quad |w_2(z)-\rho_\pm|e^{\alpha|z|}\to 0 \text{ as } z\to \mp \infty.$$
The functions $w_i$  have  derivatives of any order which vanish at infinity exponentially fast.\end{theorem}

Our main result is the stability of these fronts for the VFP dynamics, under suitable assumptions on the initial data. To state the result, we write $f_i$, solutions of (\ref{VFP}), as
$$f_i=w_i M +h_i\ .$$
Then, the perturbation $h_i$ satisfies
\begin{equation}
\partial_{t}h_i+G_i h_i=Lh_i-F_i(h)\partial_{v_z} h_i,
\label{VFP1}
\end{equation}
where the operators $G_i$ are defined by
\begin{equation}G_ih_i= v_z\partial_z h_i-(U\ast w_j')\partial_{v_z} h_i + \Big(U\ast\partial_z\int_ {\mathbb R^3} {\rm d} v h_j(\,\cdot\, ,v, t) \Big)\beta v_z M w_i \label{GG}
\end{equation}
while the force $F_i(h)$ due to the perturbation is
\begin{equation} F_i(h)=-\partial_z\int_{\mathbb R}{\rm d }z'U(z-z')\int_{\mathbb R^3} {\rm d} v h_j(z',v,t).
\end{equation}

We define $(\,\cdot\, \, ,\cdot\,)$ as the $L^2$ inner product for two scalar functions (on ${\mathbb R}$ or ${\mathbb R}\times {\mathbb R}^3$ depending on the context), while $\langle \cdot,\cdot\rangle$ denotes the $L^2$ the inner product for vector-valued functions, and we denote $\|\cdot\|$ as their corresponding $L^2$ norms. Furthermore, we define the weighted $L^2$ inner products as
$$(f_i,g_i)_M=\int_{\mathbb{R}\times\mathbb{R}^3} {\rm d }z {\rm d }v \frac{1}{w_i M} f_ig_i,\quad \langle f,g\rangle_M=\sum_{i=1,2}\int_{\mathbb{R}\times\mathbb{R}^3} {\rm d }z {\rm d }v \frac{1}{w_i M} f_i g_i,$$ with corresponding weighted $L^2$ norms by $\|\cdot\|_M$. We also define the dissipation rate as
\begin{equation}
 \|g\|_D^2=\|(I-P)g\|_M^2 +\|\nabla_v(I-P)g\|_M^2,
\end{equation}
where $P$ is the $L^2$ projection on the null space of $L=\{ cM, c\in \mathbb{R}^2 \}$, for any given $t,z$. We also define the $\gamma$-weighted norms as
$$\|g\|_{M,\gamma }=\|\zg g\|_M \qquad
\|g\|_{D,\gamma }=\|\zg g\|_{D},$$
with
$$\zg=(1+|z|^2)^\gamma .$$
In the following we will also denote by $\partial h$ the couple of derivatives $\{\partial_t h,\partial_z h\}$ with the abuse of notation $\| \partial h\|^2=\|\partial_z h\|^2+\| \partial_t h\|^2$ for any of the norms appearing below.  When there is no risk of ambiguity, to make the notation shorter, for any two vectors $f=(f_1,f_2)$ and $g=(g_1,g_2)$  we will  denote by $fg$ the vector with components $(f_1g_1,f_2g_2)$.

\vskip.3cm
The following theorem will be proved in Section 4:
\begin{theorem}\label{result} We assume that $h=(h_1,h_2)$ at time zero has the following symmetry property in $(z,v)$
\begin{equation}
h_1(z,v,0)=h_2(-z,Rv,0), \quad Rv=(v_x,v_y,-v_z).\quad \label{rsymmetry}
\end{equation}
There is $\delta_0$ small enough such that:
\begin{enumerate}
\item
If $\|h(0)\|_M+\|\partial h(0)\|_M\le \delta_0$, then there is a unique global solution to (\ref{VFP1}) such that for some $K>0$
\begin{eqnarray}&&{\frac{d}{dt}}\Big(K\big(\|h(t)\|^2_M+\|\partial_t h(t)\|^2_M\big)+\|\partial_z h(t)\|^2_M\Big)\nonumber
\\&&+K\nu_0\big(\|h(t)\|^2_D+\|\partial_t h(t)\|^2_D)+\nu_0\|\partial_z h(t)\|^2_D\le 0.
\end{eqnarray}
\item
If, for $\gamma>0$ sufficiently small,
$$\|h(0)\|_{M,\gamma}+\|\partial h(0)\|_{M,\frac12+\gamma}\le \delta_0$$
then there is constant $C>0$ such that
\begin{equation}
\sup_{0\le t\le \infty} \|h(t)\|_{M,\gamma}+\hskip-2.4pt\sup_{0\le t\le \infty}\|\partial h(t)\|_{M,\frac12+\gamma}\le
C (\|h(0)\|_{M,\gamma}+\|\partial h(0)\|_{M,\frac12+\gamma}).
\end{equation}
Moreover, we have the decay estimate
\begin{equation}
\|h(t)\|^2_M+\|\partial h(t)\|^2_M\le C\big[1+\frac{t}{{2\gamma}}\big]^{-2\gamma}\Big [\|h(0)\|_{M,\gamma}^2+\|\partial h(0)\|_{M,\frac12+\gamma}^2\Big].
\end{equation}
\end{enumerate}
\end{theorem}

A key remark to prove Theorem \ref{result} is that, since the equation preserves the symmetry property (\ref{rsymmetry}), the perturbations  $h_i(z,v,t)$ have the same simmetry property (\ref{rsymmetry}) at any time. The proof of the theorem is based on energy estimates and takes advantage of the fact that at time zero the perturbation is small in a norm involving also the space and the time derivatives. To close the energy estimates, we use the spectral gap for the Fokker-Planck operator $L$ to control $(I-P) h$, the part of $h$ orthogonal to the null space of $L$, and the conservation laws  to control $Ph$, the component of $h$ in the null space of $L$, in terms of $(I-P) h$, like the method used in \cite{Guo}.

The main difficulty in our context is the control of the hydrodynamic part $Ph$ (which can be written as $Ph=a_hM$ for some $a_h(z,t)\in \mathbb{R}^2$), in the presence  of the Vlasov force with large amplitude. Because of  the Vlasov force, the hydrodynamic equations do not give directly the control of the norm of $P h$ but instead of a norm involving the operator $A$, the second variation of the free energy $\hat{\mathcal F}$ at the front $w$, which is given, for any $g=(g_1,g_2)$ by
$$\langle g,Ag\rangle:= \sum_{i=1}^2\int_\mathbb{R} {\rm d }zg_i(z) (Ag)_i(z)=\frac{d^2}{ds^2}\hat{\mathcal F}(w+ sg)\big |_{s=0}\ .$$
The action of the operator  $A$ on $g$ is
\begin{equation}(Ag)_1=\frac{g_1}{w_1}+\beta U*g_2, \quad (Ag)_2=\frac{g_2}{w_2}+\beta U*g_1\ .\label{opa}\end{equation}

Since $w$ is a minimizer of $\hat{\mathcal F}$ the quadratic form on the left hand side  is non negative and the vanishing of the first variation of $\hat{\mathcal F}$ gives the Euler-Lagrange equations
 $$\frac{\delta \hat{\mathcal F}}{\delta \rho_i}(w)= \log w_i+\beta U*w_j-C_i=0, \quad i=1,2\ .$$
 Differentiating with respect to $z$ and using the prime to denote the derivative with respect to the $z$ variable, it results
 $$ (A w')_i=\frac{w'_1}{w_1}+\beta U*w'_j=0\ ,$$
 which shows that $w'$ is in the null space of $A$. Indeed, one can show (see Section 2) that $w'$ spans the null space of $A$ and that there exists a constant $\lambda>0$ (spectral gap)  such that
 $$\langle g,Ag\rangle\ge \lambda\sum_{i=1}^2\int_\mathbb{R} {\rm d }z\frac{1}{w_i}|(I-\mathcal P) g_i|^2$$
 where $\mathcal P$ is the projector on the null space of $A$.

 Hence, by getting estimates on the norms of $Aa_h$ and using  the spectral gap for $A$,  we can bound the component of $P h$ on the orthogonal to the null space of $A$. The component on the null space of $A$ is still not controlled. Let us write   $a_h= \alpha w'+ (I-\mathcal P) a_h$. What is missing at this stage is an estimate for $\alpha(t)=\la a_h(\,\cdot\, , t), w'\ra $ for large  times. We would like to show that $\alpha(t)$ vanishes asymptotically in time, which amounts to prove that the solution of the Vlasov-Fokker-Plank equations (VFP) converges to the initial front. The existence of a Liapunov functional for this dynamics forces the system to relax to one of the stationary points for the functional, which are of the form $M w^x$, with $w^x$  any translate by $x$ of the symmetric  front $w$. Then, it is the conservation law, in the form
 $$\int_{\mathbb{R}\times\mathbb{R}^3} {\rm d }z {\rm d }v [f(z,v,t)-M(v)w(z)]=0 $$
 which should select the front the solution has to converge to. But this is a condition requiring the control of the $L^1$ norm of the solution while our energy estimates control some weighted $L^2$ norm. In the approach in \cite{Guo} the conservation law is used in problems in finite domains or in infinite domains but in dimension greater
 or equal than $3$. The problem we are facing here is analogous to the one in \cite{CCO} and we refer to it for more discussion. One can realize the connections between the problem discussed here and the one in \cite{CCO} by looking at the hydrodynamic limit of the model.
 In \cite{MM} it is proved that  the diffusive limit  of the VFP dynamics is
\begin{equation}
\partial_{ t} \bar {\rho}=\nabla\cdot \Bigg({\mathcal M}\nabla \frac{\delta{\mathcal
F}}{\delta\bar {\rho}
}\Bigg),\quad {\mathcal M}=\beta^{-1}
\begin{pmatrix}\rho_1&0\cr 0&\rho_2\cr\end{pmatrix}
\label{0.5}
\end{equation}
where $\bar {\rho}=(\rho_1,\rho_2)$, $\displaystyle{\frac{\delta{\mathcal F}}{\delta \bar\rho}}$
denotes the
functional derivative of
${\mathcal F}$ with respect to $\bar\rho=(\rho_1,\rho_2)$ and
${\mathcal M}$ is the $2\times 2$ mobility matrix.
These equations are in the form of a gradient flow for  the  free energy functional  as the equation considered in  \cite{CCO}, which is an equation for a bounded magnetization $m(x,t)\in [-1,1]$:
$$\partial_t m=\nabla\cdot\left[\sigma(m) \frac{\delta\mathcal F}{\delta m}\right]$$
where $\s(m)=\beta(1-m^2)$ and $\mathcal F$ is a suitable non local free energy functional. In \cite{CCO} the stability result is obtained by using suitable weighted $L^2$ norms, with a weight $|x|$, which allow to control the tails of the distribution and hence a control of the $L^1$ norm.  This is possible essentially because the equation is of diffusive type.

 Unfortunately, we cannot use directly the approach in \cite{CCO}
 since the dissipation in the kinetic model is given by the Fokker-Plank operator and  does not produce directly diffusion on the space variable. In fact, we are able  to use, as explained above, $\gamma$-weighted norms (in space) with a weight $z_\gamma$, with $\gamma$ small, which are not enough to control the $L^1$ norm. Hence, to overcome the difficulty, we consider a special initial datum. We assume, as explained before, that $h$  at initial time has the particular symmetry property (\ref{rsymmetry}). It is easy to see that this property is conserved by the dynamics so that $h$ is symmetric at any later time. We note that also $w M$ is symmetric while  $w'$ is antisymmetric in the $z$ variable. This implies the vanishing at time $t=0$ of $\sum_{i=1}^2\int_{\mathbb{R}\times\mathbb{R}^3} {\rm d }z {\rm d }v h_i(z,v,t)M(v)w'_i(z)$, the component of $a$ on the null space of $A$, which consequently is zero at any later time.

Even with such a symmetry assumption (\ref{rsymmetry}), the estimate for the hydrodynamic part $Ph$ is delicate. Based on the precise spectral information of $A$, we need to further study the derivative of $A$,  $(Ag)'=\displaystyle{\frac{\partial}{{\partial z}}Ag}$. To this end, we employ the decompostion (\ref{decom}) for {\it each} component of $g$ and a contradiction argument to establish an important lower bound for $(Ag)'$, (Theorem \ref{spA'}). Furthermore, in order to get the  time decay rate, we use the additional polynomial weight function $\zg$ and a trick of interpolation to carefully derive the corresponding energy estimate in a bootstrap fashion. Once again, Theorem \ref{spA'} and its corollary (Lemma \ref{ubA}) are crucial to control local $L^2$ norm of $Ph$ in terms of its $z$-derivative.

 It is worth to stress that our result does not rely on  a smallness assumption on the potential, like for example in \cite{Vi}, where it is proved the stability in $L^1$ of the constant stationary state for a one component VFP equation, on a torus, for general initial data. The assumption of small $U$ in \cite{Vi} guarantees the uniqueness of the stationary state,
 namely it  means not to be in the phase transition region. On the contrary, we are working with values of the parameters (temperature and asymptotic values of densities, $\rho^\pm$) in the phase transition region. For values of the parameters $\rho^+=\rho^-, \ 2\beta \rho^+\le 2$ the minimizer is unique  and we can prove   that the constant solution is stable, by a simplified version of the proof given here. The critical value $\beta\rho^+=2$ is selected by the fact that the analogous of the operator $A$, that comes out from the linearization around the constant solution,  is positive and has spectral gap for $2\beta\rho^+<2$ (it coincides with the operator called $L_0$ in Theorem \ref{spA}).   We expect also that the constant solution will become unstable above this critical value.

 Finally, we want to return to the kinetic model proposed in \cite{BL}, mentioned at the beginning of this section and studied in a series of papers \cite{BELM}, in which the Fokker-Planck term is replaced by a Boltzmann kernel to model species-blind collisions between the particles. The dynamics is described by a set of two  Vlasov-Boltzmann equations, coupled through the Boltzmann collisions and the Vlasov terms and conserve not only the total masses but also energy and momentum. The stationary solutions are the same as in  the previous model, Maxwellians times densities $\rho_i$ satisfying (\ref{statio}), so that one could study the stability of these solutions with respect to the
 Vlasov-Boltzmann dynamics. This result is more difficult to get due to the non linearity of the Boltzmann terms. The first results on the stability of the Maxwellian are proved in \cite{Uk}, \cite{Ma}. Recently, it has been proved by energy methods in a finite domain or in $\mathbb R^3$ in \cite{Guo} (\cite{SG} for soft potentials) who has also extended the method to cover other models involving self-consistent forces and singular potentials \cite{Guo1}  and in $\mathbb R^d$ in \cite{LY}, who also proved the stability of a $1-d$ shock.  The stability of the non homogeneous solution for a Boltzmann equation with a given small potential force has been proved in \cite{UYZ}. We are not aware of analogous results for non small force, but a very recent one, \cite{As} relying on  the assumption that the potential is compactly supported in $\mathbb{R}^3$. Our method is in principle suited to prove stability under Vlasov-Boltzmann dynamics  on a finite interval, but what is still lacking is a detailed study of stationary solutions in a bounded domain. We plan to report on that in the future.

   The paper is organized as follows. In Section 2 we collect the properties of the operators $L$ and $A$ and the properties of the fronts. In Section 3 we prove some Lemmas that allow partial control of $Ph$ and some $z$-derivative of $Ph$ in terms of $(I-P)h$. In Section 4 we give the energy estimates for the function, the time derivative and   the $z$-derivative, which imply stability and decay of the solution.

\section{Spectral Gaps of $L$ and $A$}

In this section we collect all the relevant properties of the operators $L$ and $A$ and also the properties of the fronts.

\begin{lm}
There is  $\nu_0>0$ such that for all $g=(g_1,g_2)$,
\begin{equation}
\langle g,Lg\rangle_M\le -\nu_0\|(I-P)g\|_D^2.
\end{equation}
\label{lgap}
\end{lm}
{\bf Proof.} Since $w_i$ is bounded from below for $i=1,2$, we only need to consider the case when $g$ is a scalar.

Recall
 (\ref{fp}),  the null space of
  $L$ is clearly made of constants (in $v$) times $M$. Moreover,  $L$ is symmetric with respect to the inner product $(\,\cdot\, ,\,\cdot\,)_M$, so that $L g$ is orthogonal to the null space of $L$. We denote by $P$ the projector on the null space of $L$.
Finally, the spectral gap property holds \cite{LB}: for any $g$ in the domain of $L$
$$ \left( g, L g)_M\le -\nu((I-P) g,(I-P) g\right)_M$$
On the other hand, a direct computation yields
$$(g,Lg)_M=-\int_{\mathbb R^3} {\rm d} v\
M^{-1}\ \left|\nabla_v (I-P)g\right|^2+3\beta \int_{\mathbb R^3} {\rm d} v\
M^{-1}\ \left|(I-P)g\right|^2.$$
We thus conclude our lemma by splitting $(Lg,g)_M=(1-\epsilon)(Lg,g)_M+\epsilon(Lg,g)_M$ and applying the spectral gap property and the previous identity, for $\epsilon$ sufficiently small. \qed

\medskip

By (\ref{opa}), it is immediate to check that
$$\hat{\mathcal F}(w+\epsilon u)-\hat {\mathcal F}(w)=\epsilon^2\langle Au,u\rangle+o(\epsilon^2).$$

\begin{theorem}
There exist $\nu>0$ such that
$$\langle u, A u\rangle\ge \nu\langle{(I-\mathcal P)}u,(I-\mathcal P)u\rangle,$$
where  ${\mathcal P}$ is the  projector on ${\rm Null}\, A$:
$${\rm Null}\,A=\{ u\in L^2(\R)\times L^2(\R) \, |\,  u=c w' , c \in \R \}.$$

\label{spA}
\end{theorem}
{\bf Proof.} We first characterize ${\rm Null}\, A$. We note that (\ref{e-l'}) imply
$$\frac{u_1^2}{w_1}= - \left(\frac{u_1}{w_1'}\right)^2 \beta w'_1U*w_2' ,\quad \frac{u_2^2}{w_2}= - \left(\frac{u_2}{w_2'}\right)^2 \beta w'_2U*w_1'.$$
From (\ref{opa}), $(Au,u)$ takes the from
\begin{eqnarray}
\int_{\R}{\rm d}z \left[\frac{u_1^2(z)}{w_1(z)}+\frac{u_2^2(z)}{w_2(z)}\right] +2\beta\int_{\R}{\rm d}z\int_{\R}{\rm d}z'u_1(z)u_2(z')U(z-z') &=&\nonumber\\
-\beta\int_{\R}{\rm d}z\int_{\R}{\rm d}z' \left[\frac{u_1(z)}{w'_1(z)} - \frac{u_2(z')}{w'_2(z')}\right]^2 U(z-z')w'_1(z)w'_2(z')& . &\end{eqnarray}
But, by the monotonicity properties of $w_i$ it follows that  $-w'_1(z)w'_2(z'){\rm d}z {\rm d}z'$ is a positive measure on $\R\times \R$. Therefore the quadratic form $ $ is non negative and vanishes if and only if $h$ is parallel to $w'$.
In particular, this identifies the null space of the operator $A$.

To establish the spectral gap of $A$, it is sufficient to prove the lower bound for the normalized operator $\tilde A$: $L^2(\R)\times L^2(\R)\to L^2(\R)\times L^2(\R)$ such that
$$(\tilde A u)_i=\sqrt{w_i}(A (u\sqrt{w}))_i\ .$$
The explicit form is
$$(\tilde A u)_1= u_1 +\beta\sqrt{w_1}U*(\sqrt{w_2}u_2), \quad (\tilde A u)_2= u_2 +\beta\sqrt{w_2}U*(\sqrt{w_1}u_1)\ .$$
The corresponding associated quadratic form is
$$\langle u,\tilde A u\rangle=\int_\R {\rm d}z( u_1^2+ u_2^2)+2\beta\int_\R{\rm d}z\sqrt{w_1} u_1 U* ( u_2\sqrt{w_2})\ . $$
The operator $\tilde A$ is a bounded symmetric operator on $\mathcal H=L^2(\R)\times L^2(\R)$. From the previous considerations it is also non negative and positive on the orthogonal complement of its null space. The spectral gap for $\tilde A$ is established in \cite{CCELM3}. For completeness, we give a sketch of the proof below.

We decompose the operator as $\tilde A=\tilde A^0+ K$ where
$$(\tilde A^0 u)_1=u_1 +\beta\sqrt{\rho^+\rho^-}\ U* u_2,\quad (\tilde A^0 u)_2=u_2 +\beta\sqrt{\rho^+\rho^-}\ U* u_1,$$
\begin{eqnarray}(K u)_1&=&\beta\sqrt{w_1}U\ast (\sqrt{w_2}u_2)
-\beta\sqrt{\rho^+\rho^-}\ U\ast u_2
\\ \nonumber&=&\beta\int_{\mathbb{R}}  {\rm d} z'\left[\sqrt{w_1}(z)\sqrt{w_2}(z')-\sqrt{\rho^+\rho^-}\right] U(|z-z'|) u_2(z'),
\end{eqnarray}
\begin{eqnarray}(K u)_2&=&\beta\sqrt{w_2}U\ast (\sqrt{w_1}u_1)
-\beta\sqrt{\rho^+\rho^-}U\ast u_1\\ \nonumber
&=&\beta\int_{\mathbb{R}}  {\rm d} z'\left[\sqrt{w_2}(z)\sqrt{w_1}(z')-\sqrt{\rho^+\rho^-}\right] U(|z-z'|) u_1(z').
\end{eqnarray}
\medskip
The operator $\tilde A^0$ has the spectral gap property. Indeed, consider the equation
\begin{equation} \label{resolv} \tilde A^0 u=\lambda u+ f.\end{equation}
Denote by $\tilde u(\xi)$, $\tilde f(\xi)$  and  $\tilde U(\xi)$ the Fourier transforms of $u$, $f$ and  $U$.
We note that $\lambda$ is in the resolvent set of $\tilde A^0$ if we can find a unique solution to (\ref{resolv}), i.e. if the determinant of the matrix
$$\left (\begin{array}{cc} {1-\lambda} &{\beta}\tilde U \sqrt{\rho^+\rho^-}\\ {\beta}\tilde U\sqrt{\rho^+\rho^-}&{1-\lambda} \\\end{array}\right )$$
is different from zero for any $\xi\in R$. This happens if $\lambda$ is such that for all $\xi\in R$
$$ (1-\lambda)^2-\beta^2(\tilde U(\xi))^2 {\rho^+\rho^-}\neq 0.$$
Moreover, by the positivity of $U$, $|\tilde U(\xi)|\le\tilde U(0)=1$. As a consequence, the spectrum of $\tilde A^0$ is in the interval
$$[1- \beta  \sqrt{\rho^+\rho^-},1+ \beta\sqrt{\rho^+\rho^-}].$$ 
Now, for $\beta>\beta_c$ it is immediate to check that $\beta   \sqrt{\rho^+\rho^-}<1$ and hence the spectrum is contained in $(k,+\infty)$ for some positive $k$.

We claim that $K$ is compact on $\mathcal H$.  Indeed, uniformly for $\|u\| \le 1$, $K$ satisfies
\begin{enumerate}
\item
$\forall \epsilon >0 \quad \exists Z_\epsilon>0$:
$$\int_{|z|>Z}{\rm d} z|Ku|^2<\epsilon, \quad Z>Z_\e$$
\item
$\forall \epsilon>0\quad  \exists \ell_\epsilon>0$:
$$\int_{|z|>Z}{\rm d} z|Ku(z+\ell)-Ku(z)|^2<\e, \quad \ell>\ell_\e\ .$$
\end{enumerate}
\noindent The proof follows trivially from the regularity of the convolution, the fact that $U$ has compact support  and the fact that $\displaystyle{\lim_{x,y\to(\pm\infty,\pm\infty)}}\sqrt{w_1(x)w_2(y)}=\sqrt{\rho^+\rho^-}$. For the  property 2 the boundedness of   $w_i'$ and the regularity of $U$ are used.
Hence, by Weyl's theorem we have that the spectral gap holds also for $\tilde A$.\qed

\medskip
We are also interested into a lower bound on the norm of $(Au)'$. To this purpose, consider  $u=(u_1,u_2)\in L^2({\mathbb R})\times L^2({\mathbb R})$ with derivative $u'\in L^2({\mathbb R})\times L^2({\mathbb R})$. Assume $u$ orthogonal to $w'=(w'_1,w'_2)$: $\langle u,w'\rangle=0$.

  We now take the orthogonal decomposition of {\it each} component of $u$ with respect to the corresponding component of $w'=(w'_1,w'_2)$ in the scalar $L^2$ inner product. In terms of the vector inner product, by a direct computation, such a process leads to
\begin{equation}u=\alpha \tilde w' +\tilde u\label{decom}\end{equation}
where $\tilde u$ is such that 
$$\int_{\mathbb{R}} {\rm d} z\  \tilde u_1  w'_1=0=\int_{\mathbb{R}} {\rm d} z \ \tilde u_2  w'_2$$ 
while $\tilde w'=(w'_1,-w'_2)$ is orthogonal to $w'$ in the inner product $\langle\cdot\, ,\,\cdot\,\rangle$ (note that $w'_2(z)=-w'_1(-z)$) with the coefficient $\alpha$ computed as
$$\alpha=\frac{\langle u\, ,\,\tilde w'\,\rangle}{N},$$
$N=\langle\tilde w'\, ,\,\tilde w'\,\rangle=2\int {\rm d} z (w'_1)^2=2\int {\rm d} z (w'_2)^2$. We first prove a Lemma for $\tilde u$.

\begin{lm}
\noindent There is a constant $C$ such that
\begin{equation}
\|(A \tilde u)'\|^2\ge C \|{\mathcal Q}\tilde u'\|^2\ .
\label{sge}
\end{equation}
where
 ${\mathcal Q}$ is the orthogonal projection on the orthogonal complement of $w''$.
\label{lemma}
\end{lm}

\noindent {\bf Proof.} We follow the proof in  [CCO].
We have
$$(A \tilde u)'_i=\frac{d}{{ d} z}[\frac {\tilde u_i}{w_i}+U\ast u_j]=[\frac {\tilde u'_i}{w_i}+U\ast \tilde u'_j]-\frac{w'_i}{w_i^2}u_i= (A\tilde u')_i- \frac{w'_i}{w_i^2}\tilde u_i\ .$$
By integrating over $\bar z$ after multiplication by $ w'_i(\bar z)$  the identity
$$\tilde u_i(z)=\tilde u_i(\bar z)+\int^{z}_{\bar z}ds\ \tilde u_i'(s)$$
we get
$$\tilde u_i(z)=\frac{(-1)^{i+1}}{(\rho^+-\rho^-)}\int _{-\infty}^{+\infty} d\bar z \ w'_i(\bar z)\int^{z}_{\bar z}ds\ \tilde u_i'(s)\ ,$$
because $\int {\rm d} z \tilde u_i  w'_i=0$.

From above, we can write $(A \tilde u)'_i$ in terms of  an operator $A+K$ acting on $L^2({\mathbb R})\times L^2({\mathbb R})$ such that
$$(A \tilde u)'_i=(A\tilde u')_i+(K\tilde u')_i\ ,$$
$$(Kh)_i(z):=\frac{(-1)^i}{(\rho^+-\rho^-)}\frac{w'_i}{w_i^2}\int_{-\infty}^{+\infty}{\rm d} z'{w_i}'(z')\int^{z}_{z'}ds\  h_i(s)\ .$$
\medskip
We prove first  that

{\it The operator $K$ is compact on $L^2$.}

\medskip
 Indeed, we show that
\begin{itemize}
\item  $\forall \e>0 \quad \exists Z_\e>0$:
$$\int_{|z|>Z}{\rm d} z|K_{\bar z}h|^2<\e, \quad Z>Z_\e\ ,$$
\item   $\forall \e>0\quad  \exists \ell_\e>0$:
$$|Kh(z+\ell)-Kh(z)|^2<\e, \quad \ell<\ell_\e\ .$$
\end{itemize}
The second is true because of the continuity of the integral.  To prove the first, note that
\begin{equation}
\left|\int_{-\infty}^{+\infty}{\rm d} z'{w_i}'(z')\hskip-2.5pt\int^{z}_{z'}ds  h_i(s)\right|\hskip-3pt\le   \|h\|\int_{-\infty}^{+\infty} {\rm d} z|w_i'(z')|\sqrt{|z-z'|}\hskip-3pt\le \hskip-1.5ptC(1+|z|)\|h\|\label{puntuale}
\end{equation}
so that
\begin{eqnarray*}&&
\int_{|z|>Z}{\rm d} z\left|\frac{w'_i}{w_i^2}(z)\right|^2\left|\int_{-\infty}^{+\infty}{\rm d} z'{w_i}'(z')\int^{z}_{z'}ds\  h_i(s)\ \right|^2\\&\le& C\|h\| \int_{|z|>Z}{\rm d} z\left|\frac{w'_i}{w_i^2}(z)\right|^2|(1+|z|)^2\ .
\end{eqnarray*}
Then, by the rapid decay property of $w'_i$,
$$\int_{|z|>Z}{\rm d} z|(Kh)_i|^2\to 0, \quad Z\to +\infty,$$
which  gives the result.
Now,
$$\int_{\mathbb{R}}{\rm d} z|(A \tilde u)'|^2 =\int_{\mathbb{R}} {\rm d} z \tilde u'\left(A^2+K^*A+AK^*+K^*K\right)\tilde u' \ .$$
The operator $K^*A+AK^*+K^*K$ is compact because $A$ is bounded and $K$ compact and its null space is spanned by $w''$, because
by definition of $A+K$
$$0=(Aw')'=(A+K)w''\ .$$
But, $A^2$ has a strictly positive essential spectrum, hence
the result follows from  Weyl's theorem.
Moreover
$$\int_{\mathbb{R}} {\rm d} z|(A\tilde w')'|^2= \delta >0 ,$$
because $\tilde w'$ is orthogonal to  the null space of $A$. \qed

\medskip

\begin{theorem}\label{spgapa'}
For any $u\in L^2({\mathbb R})\times L^2({\mathbb R}), u'\in L^2({\mathbb R})\times L^2({\mathbb R})$   such that
$\left \langle u,w'\right\rangle=0$,   there exists a positive constant $B$  such that
\begin{equation}
 \|(A u)'\|^2\ge B (|\alpha |^2+\|{\mathcal Q}\tilde u'\|^2).\label{sg}
\end{equation}
where
 ${\mathcal Q}$ is the projection on the orthogonal complement of $w''$.
Furthermore, if $u'={\cal Q}u'$, then \begin{equation}
 \|(A u)'\|^2\ge B(|\alpha |^2+\|\tilde u'\|^2).\label{sgderiv}
\end{equation}
\label{spA'}
\end{theorem}
\noindent{\it\bf Proof}.
 First, we prove that there is a constant $C$ such that, if $u= {(1-\mathcal P)}u $,
\begin{equation}
\|(A u)'\|^2\ge C(\delta \alpha^2+\|(A  \tilde u)'\|^2)\ .
\label{sg1}
\end{equation}
We introduce the normalized vector $\omega$ and its decomposition along $w'$ and the orthogonal complement by setting:
$$\displaystyle{\omega=\frac {u}{\delta \alpha^2+\|(A  \tilde u)'\|^2}};\qquad \omega=\eta \tilde w'+\tilde \omega\ ,$$ 
so that  equation (\ref{sg1}) reads as
\begin{equation}\|(A  \omega)'\|^2\ge C.\ \label{sg2}\end{equation}
By the decomposition of $\omega$ we have
$$\|(A \omega)'\|^2= \|(A\tilde \omega)'\|^2 +\delta \eta^2 +2\left((A \tilde \omega)',\eta (A\tilde w')'\right)\ .$$
By definition,  $\omega$ is such that
$$\|(A\tilde \omega)'\|^2 +\delta \eta^2=1,$$
hence
$$ \|(A  \omega)'\|^2=1+2\langle(A\tilde \omega)',\eta (A\tilde w')'\rangle\ .$$

Suppose now that the inequality (\ref{sg2}) is not true. Then, for any $n$ we can find $\tilde \omega_n$ and $\eta_n$ such that
$$\|(A  [\tilde \omega_n+\eta_n\tilde w'])'\|^2=1+2\langle(A\tilde \omega)'_n, \eta_n(A\tilde w')'\rangle<\frac{1}{n}\ .$$
By weak compactness, up to subsequences, there are $\tilde \omega_0$ and $\eta_0$ such that $\tilde\omega_n$ converges weakly to $\tilde \omega_0$, $\eta_n\to \eta_0$. By  weak convergence,
$$\langle(A\tilde \omega_n)', \eta_n(A\tilde w')'\rangle\to\langle(A\tilde \omega_0)', \eta_0(A\tilde w')'\rangle$$
and
$$\text{\rm liminf} [\|(A\tilde \omega_n)'\|^2+\delta\eta_n^2]+2\langle(A\tilde \omega_0)', \eta_0(A\tilde w')'\rangle=0$$
By lower semicontinuity,
$$\|(A\tilde \omega_0)'\|^2+\delta\eta_0^2\ \le \text{\rm liminf}\left [\|(A\tilde \omega_n)'\|^2+\delta\eta_n^2\right]=1$$
Hence,
\begin{eqnarray}
0&\le&\|(A\omega_0)'\|^2=\|(A \tilde \omega_0)'\|^2 +\delta \eta_0^2+ 2\langle(A\tilde \omega_0)', \eta_0(A\tilde w')'\rangle\nonumber\\&\le& 1+2\langle(A\tilde \omega_0)', \eta_0(A\tilde w')'\rangle\le 0\ .\label{contradd}\end{eqnarray}
As a consequence,
$$\|(A  \omega_0)'\|^2=0$$
which implies $\omega_0=0$: indeed $\langle \omega_0,w'\rangle=\lim_{n\to \infty}\langle\omega_n,w'\rangle=0$ because $\omega_n$ is a sequence of vectors orthogonal to $w'$. Furthermore, since $\omega_n\to \omega_0=0$ weakly, $\eta_n\to \eta_0=0$. Then, $\langle (A\tilde \omega_0)', \eta_0(A+\tilde w')'\rangle=0$ in contradiction with last inequality in (\ref{contradd}). Therefore (\ref{sg2}) is true and, together with (\ref{sge}), implies (\ref{sg}).

\noindent Finally, to prove (\ref{sgderiv}), we notice that if $u'={\cal Q}u'$, then by (\ref{decom}), 
$$\alpha \tilde w''+\tilde u'=u'=
{\cal Q}u'=\alpha {\cal Q}\tilde w''+{\cal Q}\tilde u'.$$
We now show that ${\cal Q}\tilde w''=\tilde w''$, so that the previous identity implies ${\cal Q}\tilde u'=\tilde u'$ and hence the result. We have that
$$\langle\tilde w'',w''\rangle=(w''_1,w''_1)-(w''_2,w''_2)=0$$
because $w''_2(z)=w''_1(-z)$.\
\qed

\vskip.3cm

We conclude this Section with a pointwise bound following from previous theorem:

\begin{lm}\label{ubA}
For any function $u=(u_1,u_2)\in L^2(\mathbb{R})\times L^2(\mathbb{R})$ such that $\la u, w'\ra=0$ and $\la u,w''\ra=0$, 
there is a constant $C$ such that, for any $z\in \mathbb{R}$
\begin{equation}\label{pointwise0}|u(z)|\le C(1+|z|)\|(A u)'\|\ .
\end{equation}
\end{lm}
\vskip .3cm
\noindent {\bf Proof}.:
By using the decomposition (\ref{decom}),  
we write, using the notation of Theorem \ref{spA'},
$$u=\alpha\tilde w' +\tilde u$$
Then the argument leading to (\ref{puntuale})  provides the estimate
$$|\tilde u(z)|\le \int_{-\infty}^{+\infty}d\bar z w'(\bar z)\int_{\bar z}^z dy |\tilde u'(y)|\le (1+|z|)\|\tilde u'\| \le (1+|z|)\|(A u)'\|\ ,$$
where the second inequality uses the fast decay of $w'$ and the third one Theorem \ref{spA'}.
\noindent 
By the same theorem  we have also 
$$|\alpha|\le \|(A u)'\|.$$ 
Since $w'$ decays, we obtain (\ref{pointwise0}). \qed

\bigskip

\section{Estimates of the hydrodynamic part $Ph$.}

\vskip.3cm

We decompose the solution of (\ref{VFP1}) in the component in the null space of $L$ and in the one orthogonal to the null space:
 $h_i=P h_i+(I-P) h_i$.
We denote by $Ma_i$ the components in the null space of $L$:
 $P h_i=M\int_{\mathbb{R}^3} {\rm d} v h_i =Ma_i $, so that
$$h_i=a_iM+(I-P)h_i\ .$$ 
Since the force $F(h)$ only depends on $a$, the abuse of notation
\[F_i(a)=-\partial_z U*a_j, \quad i=1,2\ \]
will be used when convenient instead of $F_i(h)$. 
By using this decomposition  in (\ref{VFP1}) we have
\begin{eqnarray}
&\displaystyle{M\left [\partial_{t}a_i+v_z\partial_za_i-a_i U\ast w'_j M^{-1}\partial_{v_z} M +\beta v_z w_i U\ast \partial_z a_j\right]} \nonumber \\&=\displaystyle{-\partial_t(I-P)h_i-G_i (I-P)h_i- F_i(a)\partial_{v_z} h_i+ L(I-P)h_i}\label{proj},\end{eqnarray}
with $G_i$ defined in (\ref{GG}), which we rewrite for reader's convenience:
\begin{equation}G_ih_i= v_z\partial_z h_i-(U\ast w_j')\partial_{v_z} h_i + \Big(U\ast\partial_z\int_ {\mathbb R^3} {\rm d} v h_j(\,\cdot\, ,v, t) \Big)\beta v_z M w_i \label{GG1}
\end{equation}
We define
$$\mu_i=\frac{a_i}{w_i}+\beta U\ast a_j:=(A a)_i$$
so that
$$\partial_z\mu_i= \frac{1}{w_i} \partial_z a_i-a_i
\frac{w'_i}{w_i^2}+\beta U\ast\partial_z a_j$$
By using the equation for the front  (\ref{e-l'}) we  can write the equation (\ref{proj}) as
\begin{eqnarray}
&\displaystyle{M\big[\partial_{t}a_i+v_zw_i\partial_z\mu_i\big]}\label{FVP1}\\&=\displaystyle{-\partial_t(I-P)h_i-G_i (I-P)h_i- F_i(a)\partial_{v_z} h_i+ L(I-P)h_i}.\nonumber
\end{eqnarray}
By integrating (\ref{FVP1}) over the velocity, since $\int_{\mathbb{R}^3}  {\rm d} v \partial_t(I-P)h_i=0$, we have
$$\partial_{t}a_i=-\int_{\mathbb{R}^3} {\rm d} v G_i(I-P)h_i$$
and, by the definition  (\ref{GG1}) of $G_i$,
\begin{equation}
\partial_{t}a_i=-\partial_z\int_{\mathbb{R}^3} {\rm d} v v_z(I-P)h_i
\label{a}
\end{equation}
By integrating (\ref{FVP1}) over the velocity after multiplication by $v_z$ we obtain
\begin{eqnarray*}
T w_i\partial_z \mu_i=&-&\int {\rm d} v v_z\partial_t(I-P)h_i-\int_{\mathbb{R}^3} {\rm d} v v_z G_i(I-P)h_i\\&-&\int_{\mathbb{R}^3} {\rm d} v  v_zF_i(a)\partial_{v_z} h_i+\int_{\mathbb{R}^3} {\rm d} v v_zL(I-P)h_i \ .
\end{eqnarray*}
Moreover, by integrating by parts,
\begin{eqnarray*}\int_{\mathbb{R}^3} {\rm d} v v_z G_i(I-P)h_i&=&\int_{\mathbb{R}^3} {\rm d} v  v_z^2\partial_z (I-P)h_i+  U\ast w'_j \int_{\mathbb{R}^3} {\rm d} v(I-P)h_i\nonumber\\&=&\partial_z\int_{\mathbb{R}^3} {\rm d} v v_z^2 (I-P)h_i.
\end{eqnarray*}
Hence
\begin{eqnarray}
T w_i\partial_z \mu_i=&-&\int_{\mathbb{R}^3} {\rm d} v v_z\partial_t(I-P)h_i-\partial_z\int_{\mathbb{R}^3} {\rm d} v   v_z^2 (I-P)h_i\nonumber\\&+&\int_{\mathbb{R}^3} {\rm d} v v_z L(I-P)h_i+  F_i(a) a_i \ .
\label{mu}
\end{eqnarray}
We define
$$\ell^a_i=\int_{\mathbb{R}^3} {\rm d} v v_z (I-P)h_i,\quad\hskip-3.5pt  \ell^b_i=\int_{\mathbb{R}^3} {\rm d} v  v_z^2 (I-P)h_i,\quad \hskip-3.5ptm_i=\int_{\mathbb{R}^3} {\rm d} v v_z L (I-P)h_i\ .
$$
By integrating twice by parts we get the identity:
$$m_i=-\int_{\mathbb{R}^3} {\rm d} v M\partial_{v_z}\left(\frac{(I-P)h_i}{M}\right)=\beta\int_{\mathbb{R}^3} {\rm d} v v_z(I-P)h_i=\beta\ell_i^a \ .$$
The following estimates are a simple consequence of (\ref{a}) and (\ref{mu}).
$$\|\partial_t a_i\|=\|\partial_z \ell^a_i\|$$
$$\|\partial_z \mu_i\|\le\|\partial_t \ell^a_i\|+\|\partial_z \ell^b_i\|+\|m_i\|+\|\partial_z a_i\|\ \|a_i\|$$
From the definition, we have
$$|\ell_i^a|=\left|\int_{\mathbb{R}^3} {\rm d} v v_z\sqrt{M}\frac{1}{\sqrt{M}}(I-P)h_i\right|\le
C\left[\int {\rm d} v \frac{|(I-P)h_i|^2}{{M}}\right]^{\frac{1}{2}}\ .$$
This and the fact that $w$ is bounded from above and below give
$$\|\ell_i^a\|^2\le \rho^+\int_{\mathbb{R}} {\rm d} z \frac{1}{w_i}|\ell_i^a|^2\ .$$
Hence,
$$\sum_{i =1}^2\|\ell_i^a\| \le C \|(I-P)h\|_M.$$

Motivated by Theorem \ref{hydro} below, we introduce the following decomposition:
\begin{eqnarray}
\partial_z \mu^{(1)}_i&=&-\frac{1}{Tw_i}\Bigg[\int_{\mathbb{R}^3} {\rm d} v v_z(I-P)\partial_t h_i\label{1c}- \int_{\mathbb{R}^3} {\rm d} v v_z \ \\&+&
L(I-P)h_i-  F_i(a) a_i\Bigg] 
+\partial_z\left(\frac{1}{Tw_i}\right)\int_{\mathbb{R}^3} {\rm d} v \  v_z^2 (I-P)h_i,
\nonumber\\
\partial_z \mu_i^{(2)}&=&-\partial_z\left(\frac{1}{Tw_i}\int_{\mathbb{R}^3} {\rm d} v \  v_z^2 (I-P)h_i\right)
\label{2}
\end{eqnarray}
so that $\mu_i^{(1)}+\mu_i^{(2)}=\mu_i$.
We define $a^{(k)}$, $k=1,2$, by setting  $\mu_i^{(1)}=(Aa^{(1)})_i,\ \mu_i^{(2)}=(Aa^{(2)})_i$. Since the null space of $A$ is given by $\alpha w'=(\alpha w'_1,\alpha w'_2)$ for $\alpha\in \mathbb{R}$, the equation
$$\mu_i=A g_i$$
has solutions  if and only if
$$\sum_{i=1}^2 \int_{\mathbb{R}}{\rm d} z w_i'\mu_i=0$$
and they are of the form  
$$g_i=(A^{-1}\mu)_i +\theta w'_i$$
where $(A^{-1}\mu)$ is the unique solution orthogonal to the null space of $A$ and $\theta\in\mathbb{R}$. Therefore, we need to show that $\mu^{(i)}$ are orthogonal to the null space of $A$. We shall prove this it at the end of the proof of Lemma \ref{asymmetry}. Moreover, we can always choose $\theta=0$ since $a=a^{(1)}+a^{(2)}$  and $a$ does not have component on the null space of $A$. In fact, $a$ has by assumption at time zero the same symmetry property as $w$ and it is preserved in time. This implies that at any time $a$ is orthogonal to $w'$ and hence has no component in the null space of $A$. This is one of the crucial points where we use the symmetry assumption on the initial perturbation.

\medskip

We now estimate the ${L^2}$ norm  $\|\partial_z a^{(1)}\|$. To this end, we first prove that ${\mathcal Q}\partial_za_i^{(1)}=\partial_za_i^{(1)}$ which is equivalent to show that $\sum_{i=1}^2 \int \partial_za_i^{(1)}w''_i=0$.
\vskip.3cm
\begin{lm}
\label{asymmetry}
If $h_1(z,v,t)=h_2(-z,Rv,t)$, then $\langle \partial_z a^{(k)},w''\rangle =0$, $k=1,2$.
\end{lm}
\vskip.1cm
\noindent{\bf Proof:} We notice that this property is true for $\partial_z a$ because of the simmetry properties of the solution. In fact, $\partial_za_1(z)=-\partial_za_2(-z)$ and  $w''_1(z)=w''_2(-z)$.
We are left with proving that the same symmetry property hold for each $a^{(j)}$. It is then enough to prove that for $a^{(2)}$. We have that
$$a^{(2)}=-A^{-1}\left[\frac{1}{Tw_i}\int_{\mathbb{R}^3} {\rm d} v \ v_z^2 (I-P)h\right]$$
and since $A$ does not change the symmetry properties  it is enough to prove that
$$\left(\frac{1}{w_1}\int_{\mathbb{R}^3} {\rm d} v \ v_z^2 (I-P)h_1\right)(z)=\left(\frac{1}{w_2}\int_{\mathbb{R}^3} {\rm d} v \ v_z^2 (I-P)h_2\right)(-z)\ .$$
By using the properties of $h$ we have that the left hand side is equal to
$$\frac{1}{w_1(z)}\int_{\mathbb R_+^3} {\rm d} v\ v_z^2(I-P)[h_1(v,z)+h_2(v,-z)],$$
with ${\mathbb R_+^3}$ the set of velocities with $v_z\ge 0$, and the right hand side to
$$\frac{1}{w_2(-z)}\int_{\mathbb R_+^3} {\rm d} v \ v_z^2(I-P)[h_1
(v,z)+h_2(v,-z)]\ .$$
The simmetry properties of $w_i$ imply the result. The same argument also shows that $\mu^{(2)}$ is orthogonal to the kernel of $A$ and hence $\mu^{(1)}$ has the same property. This follows from the fact that  $\mu$ is in the range of $A$ which is orthogonal to the null space of $A$ because it is a symmetric operator on $L^2$. \qed

\bigskip

\begin{theorem}
We have
$$\sum_{i=1}^2\|a_i^{(2)}\|\le C\|(I-P)h \|_M\le C \|h\|_D$$
Moreover, there is $\delta_0>0$ such that, if $\|h\|_M \le \delta_0$,
$$
\sum_{i=1}^2\|\partial_z a_i^{(1)}\|\le C \big[\|(I-P)h\|_M +\|(I-P)\partial_t h\|_M\big]
$$
\label{hydro}
\end{theorem}
\noindent{\bf Remark:} Note that the estimate of $\partial_z a_i^{(1)}$ do not involve $\partial_z (I-P)h$. This is the main reason for the decomposition (\ref{1c}), (\ref{2}).
\bigskip

\noindent{\it \bf Proof:}
From (\ref{2}), by integration over $z$, since $\mu_i\to 0$ as $z\to \pm\infty$,
$$\mu_i^{(2)}=-\frac{1}{Tw_i}\int_{\mathbb{R}^3} {\rm d} v v_z^2 (I-P)h_i=\frac{1}{Tw_i}\ell_i^b$$
which implies
$$\sum_{i=1}^2\|\mu_i^{(2)}\|\le C\|(I-P)h \|_M $$
Moreover, $\mu_i^{(2)}=(Aa^{(2)})_i$ so that we have also, by Theorem \ref{spA} and the fact that $(I-\mathcal{P})a^{(k)}=a^{(k)}$, $k=1,2$,
$$\sum_{i=1}^2\|a_i^{(2)}\|\le C\|(I-P)h \|_M\ .$$
From (\ref{1c}) we get
$$\sum_{i=1}^2\|\partial_z\mu_i^{(1)}\|\le C\|(I-P)\partial_t h \|_M+C\|(I-P)h \|_M+\sup_i\|F_i(a)\|_{L^\infty}\sum_{i=1}^2\|a_i\|\ .$$
Now, by the regularity properties of $U$,
$$
\|F_i(a)\|_{L^\infty}=\| U\ast \partial_z a_j^{(1)}+\partial_z U*a_j^{(2)}\|_{L^\infty}\le C\|\partial_z a_j^{(1)}\|+C\|a_j^{(2)}\|\ .
$$
To apply Theorem \ref{spA'} we need to show that $a$ is orthogonal to $w'$. We notice that the front $w$ is symmetric under the exchange $1\to 2$ while the derivatives $w_i'$ are antisymmetric. On the other hand, as already observed,  $a$ has at time zero the same symmetry properties as $w$ and this implies that the component of $a$ on the null space of $A$ is zero at any time.
In addition, by Lemma \ref{asymmetry}, we can apply Theorem \ref{spA'} to get
\begin{eqnarray*}&&\|\partial_za^{(1)}\|\le {C}\Big[\|(I-P)\partial_t h \|_M\hskip -.1cm+\hskip -.1cm\|(I-P)h\|_M\hskip -.1cm+\hskip -.1cm\|P h\|_M(\|\partial_z  a^{(1)}\|\hskip -.1cm+\hskip -.1cm\|a^{(2)}\|)\Big]\\
&&
\le \hskip-2ptC\hskip-2pt\left[\|(I-P)\partial_t h \|_M+\|(I-P)h\|_M +\|h\|_M\ (\|\partial_z  a^{(1)}\|+\|(I-P)h\|_M)\right].
\end{eqnarray*}
Then,  for $\delta_0$ small enough
$$
\|\partial_z a^{(1)}\|\le C\big [\|(I-P)\partial_t h \|_M+C\|(I-P)h\|_M +\delta_0\|(I-P)h\|_M\big]
$$
which proves Theorem \ref{hydro}. \qed
\bigskip

\noindent As a consequence we have also
\begin{equation}
\sum_{i=1}^2\|F(h_i)\|_{L^\infty}\le C\Big[\|(I-P)\partial_t h \|_M+\|(I-P)h \|_M\Big]\ .
\label{F}
\end{equation}
From now on we use the more explicit notation $M a_h=Ph$. Moreover we use the previous decomposition: $a_{h}=a_{h}^{(1)}+a_{h}^{(2)}$. Furthermore $\|b\|_\gamma$ will denote the $\g$-weighted $ L^2$-norm $\|\zg b\|$.

\begin{lm}
\bigskip 

\label{0hydro} 
Let $0\le\g\le 1.$
Then
\begin{equation} \|a_{h}^{(2)}\|_{\gamma } \leq C\|h\|_{D,\gamma }\ ,  \label{a2gamma}\end{equation}
Moreover, there is $\delta_0>0$ such that, if $\|h\|_M\le \delta_0$, then  for any $z\in \mathbb{R}$,
\begin{equation}\label{pointwise}
|a^{(1)}_{h}(z)|\leq (1+|z|)\left(\|h\|_{D}+\|\partial _{t}h\|_{D}\right) \,
\end{equation}
and 
\begin{equation}
\|\partial _{z}a_{h}^{(1)}\|_{\gamma } \leq C\left(\|h\|_{D,\gamma
}+\|\partial _{t}h\|_{D,\gamma }\right)\ .  \label{azgamma}
\end{equation}
Furthermore, if we have also $\gamma
\leq \frac{1}{8}$,
\begin{equation}
\int d{\rm z}\frac{z^{2}}{(1+z^{2})^{2-2\gamma }}|a_{h}|^{2} \leq
C\left(\|h\|_{D,\gamma }^{2}+\|\partial _{t}h\|_{D,\gamma }^{2}\right)\ .
\label{agamma}
\end{equation}
\end{lm}
\vskip.4cm
\noindent\textbf{Proof: }
\noindent 
We introduce the  {\it  commutator} $[\zg, A]$ defined as follows:  for any function $a=(a_1,a_2)$
\begin{equation*}
[\zg ,A]a=\zg A a-A(\zg a),
\end{equation*}
which, by the definition of $A$, up to a factor $\beta$, reduces to
$$[\zg ,U]a\equiv\zg U*  a-U*(\zg a)$$
The commutator can be estimated by taking into account the property of the convolution and the fact that $U$ is of finite range.  Indeed,  it is easy to check that, if $|z-z'|<R$, then, with the notation $\zgp=(1+z'^2)^\gamma$, we have 
\begin{equation}\label{z-z'}
|\zg-\zgp|\le \gamma (\zga+\zgap) |z-z'|\le C\gamma\zgap|z-z'|
\end{equation}
for some constant $C$. Therefore
\begin{eqnarray*}
|[\zg ,U]a(z)|&=&\Big|\int dz' U(|z-z'|)(\zg -\zgp)a(z')\Big|\\&\le& C\int dz' U(|z-z'|)|z-z'|\zgap a(z')
\end{eqnarray*}
Hence
\begin{equation}\label{stimacomm}
\|[\zg ,U]a\| \leq C\|\zga a\|\ 
\end{equation}
and in particular, 
$$
\|[\zg ,A]a_{h}^{(2)}\|\leq C\|\zga a_{h}^{(2)}\|.$$
Moreover
\begin{equation*}
\zg A a_{h}^{(2)}=-\zg \frac{1}{Tw_{i}%
}\int dvv_{z}^{2}(I-P)h_{i}.
\end{equation*}
which implies
$$\|\zg A a_{h}^{(2)}\|\le C\|h\|_{D,\gamma}\ .$$
The last two estimates together imply
$$\|A(\zg a_{h}^{(2)})\|\le C\|h\|_{D,\gamma}+\|a_{h}^{(2)}\|_{\gamma-\frac{1}{2}}\ .$$
Since $\zg a_{h}^{(2)}$ has the same symmetry properties of $a_{h}^{(2)}$, it is orthogonal to $w'$ as well and we can use Theorem \ref{spA}  to deduce
$$\|a_{h}^{(2)}\|_\g\le C(\|h\|_{D,\gamma}+\|a_{h}^{(2)}\|_{\gamma-\frac{1}{2}})\ .$$
By repeating the argument with $\gamma-\frac{1}{2}$ instead of $\gamma$  and using the fact that $z_{\g-1}<1$ for $\g\le 1$, by the first part of Theorem \ref{hydro}, we obtain (\ref{a2gamma}).

\vskip.2cm\noindent

To prove the the second statement it is enough to note that $a^{(1)}_h$ is orthogonal to $w'$ by construction and to $w''$ by symmetry (Lemma \ref{asymmetry}). Hence we have, by Lemma \ref{ubA},
\begin{equation}\label{pointwise2}|a^{(1)}_h(z)|\le C(1+|z|)\|(A a^{(1)}_h)'\| ,
\end{equation}
and, by Theorem \ref{hydro}, we obtain (\ref{pointwise}).

\vskip.3cm

To estimate $\partial _{z}a_{h}^{(1)}$ we note that  $\displaystyle{\partial _{z}(Aa_{h}^{(1)})_{i}=(A\partial _{z}a_{h}^{(1)})_{i}-%
\frac{w_{i}^{\prime }}{w_{i}^{2}}(a_{h}^{(1)})_i}$. Therefore
\begin{equation}
\zg (A\partial _{z}a_{h}^{(1)})_{i}=\zg \partial _{z}(Aa_{h}^{(1)})_{i}+\frac{\zg w_{i}^{\prime }
}{w_{i}^{2}}(a_{h}^{(1)})_i.
\label{bas}
\end{equation}
Clearly, since $w_{i}^{\prime }$ decays exponentially, by (\ref{pointwise}), we have the following estimate for the second term in (\ref{bas})
\begin{equation*}
\|\frac{\zg w_{i}^{\prime }}{w_{i}^{2}}(a_{h}^{(1)})_i\| \leq
C\big(\|h\|_{D}+\|\partial _{t}h\|_{D}\big)
\end{equation*}
We examine now the first term in (\ref{bas}). We  deduce from  equation (\ref{1c})
\begin{equation*}
\|\zg \partial _{z}(Aa_{h}^{(1)})\|\leq
C\left(\|h\|_{D,\gamma }+C\|\partial _{t}h\|_{D,\gamma}+\|\zg F(a_{h})\|_{L^\infty}\  \|a_h\|\right)\ .
\end{equation*}
We further split $\|\zg F(a_{h})\|_{L^\infty}$ in the last term as
\begin{eqnarray*}
&&\|\zg F(a_{h}^{(2)})\|_{L^\infty}+\|\zg U\ast
(\partial _{z}a_{h}^{(1)})\|_{L^\infty}\\
&
\leq&  \|F(\zg (a_{h}^{(2)}))\|_{L^\infty}+\|[\zg ,\partial_zU]a_{h}^{(2)}\|_{L^\infty}\\&+&\|U\ast (\zg \partial
_{z}a_{h}^{(1)})\|_{L^\infty}+\|[\zg ,U]\partial _{z}a_{h}^{(1)}\|_{L^\infty} \\
&\leq& \left(\|\zg a_{h}^{(2)}\|
+\|\zg \partial _{z}a_{h}^{(1)}\| \right)
\leq C\left(\|h\|_{D,\gamma }+\|\zg \partial _{z}a_{h}^{(1)}\| \right) \ .
\end{eqnarray*}
We have used (\ref{a2gamma}) in the last inequality.   The commutators are estimated as before in (\ref{stimacomm}), leading to a term
\[\|\zga \partial _{z}a_{h}^{(1)}\|+\|\zga a_{h}^{(2)}\|\le \|\zg \partial _{z}a_{h}^{(1)}\|+\|h\|_{D,\gamma }\ ,\]
where we have used again  (\ref{a2gamma}).
We hence conclude that, for $\|a_h\|$ small, 
\begin{eqnarray*}
\|A(\zg \partial _{z}a_{h}^{(1)})\|  &\leq
&\|\zg A\partial
_{z}a_{h}^{(1)}\| +\|[\zg ,A]\partial _{z}a_{h}^{(1)}\|  \nonumber\\
&\leq &C\left(\|h\|_{D,\gamma }+\|\partial _{t}h\|_{D,\gamma
}+\|\zg \partial _{z}a_{h}^{(1)}\| \ \|a_h\| \right)
\label{vaffa3}\ .
\end{eqnarray*}
We now decompose along the null space of $A$ and its orthogonal complement in order to use the spectral gap of $A$:  Denote by  
$$\tau= w'\|w'\| ^{-1}$$ 
the unit vector in the direction $w'$.

By the decay of $w'$, for any $ L^2$ function $q$, we get
 \begin{equation}\label{stimatau}
 |\la \zg q, \tau\ra |\le \|q\|\  .
 \end{equation}
 Hence we have
\begin{eqnarray*}
\|\zg \partial _{z}a_{h}^{(1)}\|  &\leq
&\|\zg \partial _{z}a_{h}^{(1)}-\langle \zg \partial _{z}a_{h}^{(1)},\tau\rangle \tau\|  +\|\langle \zg \partial _{z}a_{h}^{(1)},\tau\rangle \tau
\|  \\
&\leq &C\|A(\zg \partial
_{z}a_{h}^{(1)})\| +C\{\|h\|_{D}+\|\partial _{t}h\|_{D}\} \\
&\leq &C\{\|h\|_{D,\gamma }+\|\partial _{t}h\|_{D,\gamma
}\}+C\|\zg \partial _{z}a_{h}^{(1)}\| \  \|a_h\|
\end{eqnarray*}
Putting estimates together, for $\|a_h\| $ small,
\begin{equation}
\|\zg \partial _{z}(Aa_{h}^{(1)})\|\leq C\{\|h\|_{D,\gamma }+\|\partial _{t}h\|_{D,\gamma
}\}
\label{vaffa1}
\end{equation}
and hence we deduce (\ref{azgamma}).

To prove (\ref{agamma}), first note that the contribution due to $a_h^{(2)}$ is easily bounded by (\ref{a2gamma}). As for the contribution due to $a_{h}^{(1)}$, since we can use the pointwise estimate (\ref{pointwise}),  the key is to estimate $\displaystyle{\frac{za_{h}^{(1)}}{(1+z^{2})^{1-\gamma }}}$ for $z$ large. 

\noindent In fact, let $\chi (z)$ be a smooth
cutoff function with $\chi (z)\equiv 1$ for $|z|\geq k$, for $k$ large and $\chi (z)\equiv 0$ for $|z|\leq k-1$. We have for the contribution from $a_h^{(1)}$ due to $|z|\le k$, by (\ref{pointwise})
\begin{eqnarray*}
\int d{\rm z}\ \frac{z^{2}}{(1+z^{2})^{2-2\gamma }}&&\{1-\chi \}|a_{h}^{(1)}|^{2}\\ \leq \Big(\|h\|_{D}^{2}+\|\partial
_{t}h\|_{D}^{2}\Big)
\int_{|z|\leq k}d{\rm z}&&\frac{z^{2}(1+|z|)^2}{(1+z^{2})^{2-2\gamma }}\ \le C_k\Big(\|h\|_{D}^{2}+\|\partial
_{t}h\|_{D}^{2}\Big) \ .
\end{eqnarray*}
We now consider the contribution for $|z|\,\ $large. We have:
\begin{eqnarray*}&&
\int d{\rm z}\  \frac{\chi z^{2}|a_{h}^{(1)}|^{2}}{(1+z^{2})^{2-2\gamma }} =-\int d{\rm z}\
\frac{d}{dz}\left(\frac{1}{2(1-2\gamma )(1+z^{2})^{1-2\gamma }}\right)z\chi
|a_{h}^{(1)}|^{2} \\&&
=\hskip-4pt\int \hskip-3ptd{\rm z}\ \frac{1}{2(1-\gamma )(1+z^{2})^{1-2\gamma }}z\chi ^{\prime
}|a_{h}^{(1)}|^{2}\hskip -.1cm+\hskip -.1cm\int d{\rm z}\ \frac{1}{2(1-2\gamma )(1+z^{2})^{1-2\gamma }}\chi
|a_{h}^{(1)}|^{2} \\&&
+\int d{\rm z}\ \frac{1}{(1-2\gamma )(1+z^{2})^{1-2\gamma }}\chi a_{h}^{(1)}\partial
_{z}a_{h}^{(1)}.
\end{eqnarray*}
Therefore,
\begin{eqnarray*}
&&\int d{\rm z}\ \left(\frac{z^{2}}{(1+z^{2})^{2-2\gamma }}-\frac{1}{2(1-2\gamma
)(1+z^{2})^{1-2\gamma }}\right)\chi |a_{h}^{(1)}|^{2}= \\
&&\hskip -.2cm\int \hskip-3pt d{\rm z}\frac{1}{2(1-2\gamma )(1+z^{2})^{1-2\gamma }}z\chi ^{\prime
}|a_{h}^{(1)}|^{2}\hskip -.1cm+\hskip -.1cm\int\hskip-3pt d{\rm z}\frac{1}{(1-2\gamma )(1+z^{2})^{1-2\gamma }}\chi
a_{h}^{(1)}\partial _{z}a_{h}^{(1)}.
\end{eqnarray*}
For $\gamma $ $\leq \frac{1}{8}$ and $|z|>k$,
\begin{equation*}
\frac{z^{2}}{(1+z^{2})^{2-2\gamma }}-\frac{1}{2(1-2\gamma
)(1+z^{2})^{1-2\gamma }}\geq \frac{z^{2}}{4(1+z^{2})^{2-2\gamma }}.
\end{equation*}
Since $\chi ^{\prime }\equiv 0$ for $|z|\geq k$, by using again (\ref{pointwise}) to bound the term with $\chi'$, we obtain:
\begin{eqnarray*}
&&\int d{\rm z}\ \frac{z^{2}}{4(1+z^{2})^{2-2\gamma }}\chi |a_{h}^{(1)}|^{2} \leq
\frac{1}{8}\int d{\rm z}\ \frac{z^{2}}{
(1+z^{2})^{2-2\gamma }}\chi |a_{h}^{(1)}|^{2}\\&& +C\|\zg \partial _{z}a_{h}^{(1)}\|^{2}+ C_k\Big(\|h\|_{D}^{2}+\|\partial
_{t}h\|_{D}^{2}\Big).
\end{eqnarray*}
We thus deduce (\ref{agamma}) by using (\ref{azgamma}) and conclude the proof of Lemma \ref{0hydro}. \qed

\bigskip
It will be important in the energy estimate in next section, and in particular in the proof of Lemma \ref{zweight}, to bound $\partial _{z}(\zg \partial _{z}a^{(1)})$ in terms of at most one space derivative of $h$. To this end it is convenient to introduce the quantity $a_{h}^{(3)}$ defined by the positions:
\begin{equation}\label{defa3}
(Aa_{h}^{(3)})_i\equiv -\frac{1}{Tw_i}\int d{ v}\ v_{z}\partial _{t}(I-P)h_{i}, \quad \langle a_{h}^{(3)},w^{\prime }\rangle =0\ .
\end{equation}
We note that, by Theorem \ref{spA} and the orthogonality condition, it follows that
\begin{equation}\label{l2a3}
\|a_h^{(3)}\|\le C\|\partial_t h\|_D.
\end{equation}
We have:
\begin{lm}
\label{azz}\bigskip There is $\delta_0>0$ such that, if $\|a_{h}\| +\|\partial _{z}a_{h}\| \le\delta_0$, the following estimates 
\begin{eqnarray}
\|\partial _{z}a_{h}^{(2)}\|_\gamma &\leq &C\left(\|\partial _{z}h\|_{D,\gamma
}+\|h\|_{D}\right) \ ,  \label{ag2} \\
\| a_{h}^{(3)}\|_\gamma&\leq &C\|\partial
_{t}h\|_{D,\gamma }\   \label{ag3} \end{eqnarray}
hold  for $0\leq \gamma \leq {1}$.
Moreover, if $\|\partial _{z}a^{(1)}_{h}\|_{\frac{1}{2}} \le \eta$ for some finite constant $\eta$, then there is $C_\eta$ such that 
\begin{eqnarray}\|\partial _{z}\big[\zg \partial
_{z}a_{h}^{(1)}-\zg a_{h}^{(3)}\big]\|  &\leq &C_\eta\left(\|\partial h\|_{D,\gamma }+\|h\|_{D,\gamma -\frac{1}{2}}\right)\ .  \label{ag1}
\end{eqnarray}
\end{lm}

\noindent \textbf{Proof.} For notational simplicity, we denote 
$$a_{\g}=\zg \partial _{z}a_{h}$$ 
and similar meaning will have $a_\g^{(k)}$, $k=1,\dots,3$. We need to estimate
$\|a_{\g}^{(3)}\| $, $\|a_{\g}^{(2)}\| $ and $\|\partial
_{z}a_{\g}^{(1)}\| $.

 First of all, we prove (\ref{ag3}). From the definition of $Aa^{(3)}_h$ we have
\begin{eqnarray*}
\|A(\zg a_{h}^{(3)})]\|  &\leq &\|\zg Aa_{h}^{(3)}\| +\| [A,\zg ]a_{h}^{(3)}\|  \\
&\leq &\|\partial _{t}h\|_{D,\gamma}+C\|\zga a_{h}^{(3)}\| \\
&\leq & \|\partial _{t}h\|_{D,\gamma} +C\|a^{(3)}_h\|_{\gamma-\frac{1}{2}} \ .
\end{eqnarray*}
We then decompose $\zg a_h^{(3)}$ along the direction $\tau$  (recall that $\tau=w'\|w'\|^{-1}$) and its orthogonal complement $(\zg a_h^{(3)})^\perp$:
$$\zg a_h^{(3)}= \la \zg a_h^{(3)},\tau\ra\tau+ (\zg a_h^{(3)})^\perp.$$
We deduce, again by Theorem \ref{spA},
\begin{equation*}
\|(\zg a_{h}^{(3)})^\perp\| \leq C\|\partial _{t}h\|_{D,\gamma }+C\|a^{(3)}_h\|_{\gamma-\frac{1}{2}}\ .
\end{equation*}
 The component of $\zg a_{h}^{(3)}$ along $\tau$ can be bounded by using  (\ref{stimatau})  and (\ref{l2a3}).
The proof of  (\ref{ag3}) is completed by repeating the argument with $\g$ replaced by $\g-\frac 12$ and applying the bound to $\| a_{h}^{(3)}\|_{\gamma-\frac{1}{2}}$.

We now turn to (\ref{ag2}). Note that
\begin{eqnarray}
Aa_{\g}^{(2)} &=&A(\zg \partial _{z}a_{h}^{(2)}) \nonumber \\
&=&\zg A(\partial _{z}a_{h}^{(2)})-[\zg ,A]\partial _{z}a_{h}^{(2)}\label{zgammadza2h} \\
&=&\zg \partial _{z}Aa_{h}^{(2)}+\zg
\frac{w^{\prime }}{w^{2}}a_{h}^{(2)}-[\zg ,U]\partial _{z}a_{h}^{(2)}.\nonumber
\end{eqnarray}
Clearly, by the decay of $w'$, $\displaystyle{\|\zg \frac{w^{\prime }}{w^{2}}
a_{h}^{(2)}\|\le C\|a_h^{(2)}\| \leq C\|h\|_{D}}$ by Theorem \ref{hydro}. Since

\begin{equation}
\zg \partial _{z}(Aa_{h}^{(2)})_i=-\zg \partial _{z}\Big(\frac{1}{Tw_{i}}\int dvv_{z}^{2}(I-P)h_{i}\Big)=\label{zgammadza2hp}
\end{equation}
\[ -\zg \Big(\frac{1}{Tw_{i}}\int dvv_{z}^{2}(I-P)\partial _{z}h_{i}\Big)+\zg\frac{w'_i}{T(w_{i})^2}\int dvv_{z}^{2}(I-P)h_{i}\ ,
\] 
again by Theorem \ref{hydro} and the decay of $w'$ 
it follows that
\begin{equation}\label{zgammadza2hs}
\|\zg \partial _{z}Aa_{h}^{(2)}\| \leq C\Big(\|\partial
_{z}h\|_{D,\gamma }+\|h\|_D\Big).
\end{equation}
By (\ref{z-z'}) we have for the commutator $[\zg ,U]\partial
_{z}a_{h}^{(2)}$:
\begin{equation}\label{zgammadza2ht}
\left\|\int d{\rm z}'U(|z-z^{\prime }|)(\zg -\zgp)\partial _{z}a_{h}^{(2)}(z^{\prime })\right\| \leq
C\|\zga \partial_za_{h}^{(2)}\|.
\end{equation}
We therefore can decompose
\begin{equation*}
a_{\g}^{(2)}=\langle a_{\g}^{(2)},\tau\rangle
\tau+(a_{\g}^{(2)})^\perp.
\end{equation*}
By (\ref{stimatau}) and Theorem \ref{hydro} we have
\begin{equation*}
\|\langle a_{\g}^{(2)},\tau\rangle \tau\| \leq \|a_{h}^{(2)}\| \leq C\|h\|_{D}.
\end{equation*}
But $\|(a_{\g}^{(2)})^\perp\| $ is bounded by using the spectral gap of $A$ and the inequality
\begin{equation}
\|Aa_{\g}^{(2)}\| \leq C\left(\|h\|_{D }+\|\partial_{z}h\|_{D,\gamma }+\|\zga\partial_za_{h}^{(2)}\|\right),
\label{vaffa2}
\end{equation}
Collecting terms and iterating once the inequality, as before, we deduce (\ref{ag2}).

Finally, to estimate $\partial _{z}(a_{\g}^{(1)}-\zg a_{h}^{(3)})$,  we use the commutation relation
\begin{equation}\label{commuta}
A\partial _{z}a_{i}=\partial
_{z}(Aa)_{i}+\frac{w'_{i}}{w_{i}^{2}}a_i
\end{equation}
to get 
\begin{eqnarray*}
(Aa_{\g}^{(1)})_i &=&\left(A(\zg \partial _{z}a_{h}^{(1)})\right)_i \\
&=&\zg \partial_{z}(Aa_{h}^{(1)})_i-([\zg ,A]\partial_{z}a_{h}^{(1)})_i+\frac{\zg w_{i}^{\prime }(a_{h}^{(1)})_i}{w_{i}^{2}}
\end{eqnarray*}
By equation (\ref{1c}) and the definition of $Aa_h^{(3)}$
\begin{eqnarray*}
& &\zg\partial_{z}(Aa_{h}^{(1)})_i=\\
& &\zg(Aa^{(3)}_h)_i +\zg \left[\frac{1}{Tw_i}\left[ \int_{\mathbb{R}^3} {\rm d} v v_z
L(I-P)h_i+  F_i(h) (a_{h})_i\right]\right.\\
& &\left.+\partial_z\left(\frac{1}{Tw_i}\right)\int_{\mathbb{R}^3} {\rm d} v \  v_z^2 (I-P)h_i
\right].
\end{eqnarray*}
Therefore,
\begin{eqnarray*}
& &(Aa_{\g}^{(1)})_i-\zg(Aa^{(3)}_h)_i= \left[\frac{\zg}{Tw_i}\left[\int_{\mathbb{R}^3} {\rm d} v v_z
L(I-P)h_i+  F(h_j) (a_{h})_i\right]\right.\\
& &\left.+\partial_z\left(\frac{1}{Tw_i}\right)\int_{\mathbb{R}^3} {\rm d} v \  v_z^2 (I-P)h_i
\right]
-([\zg ,A]\partial_{z}a_{h}^{(1)})_i+\frac{\zg w_{i}^{\prime }(a_{h}^{(1)})_i}{w_{i}^{2}}.
\end{eqnarray*}
Using again the commutation relation (\ref{commuta})  and the previous relation, we find
\begin{eqnarray*}
&&\big(A\partial _{z}(a_{\g}^{(1)}-\zg a_{h}^{(3)})\big)_{i} \\
&&=\partial _{z}\big(Aa_{\g}^{(1)}-(A(\zg a_{h}^{(3)})\big)_{i}-\frac{w_{i}^{\prime }}{w_{i}^{2}}
(a_{\g}^{(1)}-\zg a_{h}^{(3)})_{i}\\
&&=\partial_z\Big[ (Aa_{\g}^{(1)})_i-\zg (Aa^{(3)}_h)_i-([\zg,A]a^{(3)}_h)_i\Big]-
\frac{w_{i}^{\prime }}{w_{i}^{2}}
(a_{\g}^{(1)}-\zg a_{h}^{(3)})_{i}
\\
&&=\partial_{z}\Bigg\{ \zg \Bigg[\frac{1}{Tw_i}\Big(\int_{\mathbb{R}^3} {\rm d} v v_z
L(I-P)h_i+  F_i(a_h)(a_h)_i\Big) \\
&&
+\partial_z\Big(\frac{1}{Tw_i}\Big)\int_{\mathbb{R}^3} {\rm d} v \  v_z^2 (I-P)h_i\Bigg]-([\zg ,A]\partial_{z}a_{h}^{(1)})_i-\frac{\zg w_{i}^{\prime }(a_{h}^{(1)})_i}{w_{i}^{2}}\\
&&-([\zg ,A]\partial _{z}a_{h}^{(1)})_i -([A,\zg ]a_{h}^{(3)})_i\Bigg\}+\frac{w_{i}^{\prime }}{w_{i}^{2}}(a_{\g}^{(1)}-\zg a_{h}^{(3)})_{i}
\end{eqnarray*}
The terms involving the commutator can be estimated by moving the $z$-derivative on the potential $U$ inside the convolution.

\noindent We only need to estimate $\displaystyle{\partial _{z}\left( \frac{\zg }{Tw_{i}}
F_i(a_h)(a_h)_i\right)}$, all the other terms being estimated by arguments already used. We expand it as
\[\partial _{z}\left(\frac{\zg }{Tw_{i}}\right) F_{i}(a_hh)( a_h)_i-
\frac{\zg }{Tw_{i}}\left[(\partial_z U\ast \partial
_{z}( a_h)_j)( a_h)_i\hskip-1.5pt-\hskip-1.5pt(U\ast \partial _{z}( a_h)_j)\partial _{z}( a_h)_i\right]\]
The first term is bounded by
\[\left(\|\zga a_{h}^{(2)}\| +\|\zga \partial _{z}a_{h}^{(1)}\| \right)\|a_{h}\| \]
We modify the last two terms above (up to the factor $(Tw_i)^{-1}$) as follows:
\begin{eqnarray*}&-&(\partial_z U\ast \zg \partial
_{z}( a_h)_j)( a_h)_i-(U\ast \zg \partial _{z}( a^{(2)}_h)_j)\partial _{z}( a_h)_i\\&-&(U\ast z_{\gamma-\frac{1}{2}}( \partial_za^{(1)}_h)_j)z_{\frac{1}{2}} \partial _{z}( a_h)_i
+[\partial_z U,\zg ]\partial
_{z}( a_h)_j( a_h)_i
\\&+&([U,\zg ]\partial
_{z}( a^{(2)}_h)_j+[U,z_{\gamma-\frac{1}{2}} ]\partial
_{z}( a^{(1)}_h)_j)\partial_z( a_h)_i\end{eqnarray*}
The $ L^2$ norms of the last two terms are bounded by
\begin{eqnarray*}\Big(\|\zga a_{h}^{(2)}\| &+&\|\zga \partial _{z}a_{h}^{(1)}\| \Big)\Big(\|a_{h}\| +\|\partial_za_{h}\| \Big)\\&&\le \delta_0\Big(\|h_{D,\gamma-\frac{1}{2}}\| +\|\partial_t h_{D,\gamma-\frac{1}{2}}\|\Big)\end{eqnarray*}
The inequality follows from Lemma  \ref{0hydro} and the smallness assumption $\|a_{h}\| +\|\partial _{z}a_{h}\| \le\delta_0$.
The contribution from $a_h^{(2)}$ to the first term is easily bounded by the first part of Lemma \ref{azz}. We write the contribution to the first term due to $a_h^{(1)}$ (up to the minus sign)  as
$$\left( U\ast \partial_z\left((a_{\g}^{(1)})_j-\zg  (a_h^{(3)})_j\right)+\partial_z U\ast \zg (a_{h}^{(3)})_j\right)( a_h)_i$$
Finally, we get
\begin{eqnarray*}
&&\left\|{\partial _{z}\left( \frac{\zg }{Tw_{i}}
F_i(a_h)( a_h)_i\right)}\right\|
\leq  \delta_0(\|h_{D,\gamma-\frac{1}{2}}\| +\|\partial_t h\|_{D,\gamma-\frac{1}{2}}\|)\\
&&\qquad+\left(\|\partial
_{z}(a_{\g}^{(1)}-\zg a_{h}^{(3)})\| +
\|\zg a_{h}^{(3)}\|+\|\zg \partial_z a_{h}^{(2)}\|\right)\\
&&\qquad\times \left(\|a_{h}\| +\|\partial _{z}a_{h}\| \right)+\|\partial_za^{(1)}\|_{\gamma-\frac{1}{2}}\|\partial_z a_{h}\|_{\frac{1}{2}}.
\end{eqnarray*}
We use (\ref{ag2}), (\ref{ag3}) and Lemma \ref{0hydro} to get
$$\|\zg a_{h}^{(3)}\|+\|\zg \partial_z a_{h}^{(2)}+\|\partial_za^{(1)}\|_{\gamma-\frac{1}{2}}\|\le \|h\|_{D,\gamma-\frac{1}{2}}+\|\partial h\|_{D,\gamma}\ .$$
We therefore conclude, by using $\|\partial_za^{(1)}\|_{\frac{1}{2}}\le C$,
\begin{eqnarray}
&&\|A(\partial _{z}a_{\g}^{(1)}-\zg a_{h}^{(3)})\|
\label{apart} \leq C\big(\|\partial h\|_{D,\gamma }\nonumber \\& &+\|h\|_{D,\gamma -\frac{1}{2}
}+\|\partial _{z}(a_{\g}^{(1)}-\zg a_{h}^{(3)}\| )
(\|a_{h}\| +\|\partial _{z}a_{h}\| ).
\end{eqnarray}
We then split $\|\partial _{z}(a_{\g}^{(1)}-\zg a_{h}^{(3)})\| $ into
\begin{eqnarray*}
&&\|\langle \partial _{z}(a_{\g}^{(1)}-\zg a_{h}^{(3)}),\tau\rangle\tau\| +\left\|\Big(\partial _{z}(a_{\g}^{(1)}-\zg a_{h}^{(3)})\Big)^\perp\right\|
\end{eqnarray*}
By using (\ref{stimatau}), Theorem \ref{hydro} and (\ref{l2a3}), the first term is bounded by $\|h\|_D$.  The second can be absorbed in the left hand side
for $\{\|a_h\|\| +\|\partial _{z}a_h\| \}$ small, by using the spectral gap for $A$. This  concludes the proof of   (\ref{ag1}). \qed
\vskip.3cm
\goodbreak

\bigskip

\section{Energy Estimates and Decay.}

In this section we obtain bounds on the $L^2$-norms of the perturbation and its space and time derivatives,
which will ensure the stability of the front solution, as well as on the $\g$-weighted norms which control the space decay of the perturbation and, as a consequence, the rate  of convergence to zero of the perturbation as $t\to\infty$. All the estimates are obtained via an energy method based on a notion of ``energy'' which is constructed in terms of the linearization of the Liapunov functional $\mathcal G$, which replaces the usual entropy functional in the case of long range interactions. The so obtained energy involves the quadratic form associated to the operator $A$, as discussed in the Introduction.  All the estimates are based on the following Lemma, depending on the structure of the linearized equation, which is common to the equation for the perturbation as well as to the one for its derivatives.

\begin{lm}
\label{identity}\bigskip Given $\G=(\G_1,\G_2)$, let $g=(g_1,g_2)$ be the solution to the equation
\begin{equation}
\partial _{t}g_{i}+G_{i}g_{i}-Lg_{i}=\Gamma _{i},  \label{model}
\end{equation}
with $G_ig_i$ defined in (\ref{GG}). Then, with the usual orthogonal decomposition
$$g=a_gM+(1-P)g,$$
we have:
\begin{eqnarray*}
\frac{1}{2}\frac{d}{dt}\Bigg\{ \int_{\mathbb{R}} \mathrm{d}z a_{g}Aa_{g} &+&\sum_{i=1}^{2}\int_\mathbb{R}\mathrm{d}z\int_{\mathbb{R}^3}\mathrm{d}v \frac{
1}{Mw_{i}}|(I-P)g_{i}|^{2}\Bigg\} \\ &-&\sum_{i=1}^{2}\int_{\mathbb{R}\times \mathbb{R}^3}\mathrm{
d}z\mathrm{
d}v\frac{1}{w_{i}M}(I-P)g_{i}L(I-P)g_{i} \\
&=&\langle Aa_{g},\Gamma \rangle +\langle
\frac{1}{Mw}(I-P)g,\Gamma\rangle .
\end{eqnarray*}
\end{lm}
Note that the inner product in right hand side is just the $L^2(dzdv)$ inner product.

\medskip
\noindent\textbf{Proof:}
Repeating the same
computation as in Section 3, we have
\begin{equation}
M\big(\partial _{t}(a_{g})_i+v_{z}w_{i}\partial _{z}(Aa_g)_{i}\big)=-\partial
_{t}(I-P)g_{i}-G_{i}(I-P)g_{i}+L(I-P)g_{i}+\Gamma _{i}.  \notag
\end{equation}
We take the scalar product $( \,\cdot \,,\,\cdot \,)_{M}$ of (\ref
{model}) with $Mw_i(Aa_g)_{i}+(I-P)g_{i}$ to get:
\begin{eqnarray*}
&&\frac{1}{2}\sum_{i=1}^{2}\frac{d}{dt}\left[ \int_{\mathbb{R}\times \mathbb{R}^{3}}\mathrm{d}z\mathrm{d}vM(a_{g})_i(Aa_g)_{i}+\int_{\mathbb{R}\times \mathbb{
R}^{3}}\mathrm{d}z\mathrm{d}v\frac{1}{Mw_{i}}|(I-P)g_{i}|^{2}\right]= 
\\
&&-\sum_{i=1}^{2}\Bigg\{\int_{\mathbb{R}\times \mathbb{R}^{3}}\hskip-5pt\mathrm{d}z\mathrm{d}
vMw_{i}(Aa_g)_{i}v_{z}\partial_{z}(Aa_g)_{i}\hskip-.1cm-\hskip-.1cm\int_{\mathbb{R}\times \mathbb{R}
^{3}}\hskip-5pt\mathrm{d}z\mathrm{d}v(I-P)g_{i}v_{z}\partial _{z}(Aa_g)_{i} \\
&&-\int_{\mathbb{R}\times \mathbb{R}^{3}}\hskip-5pt\mathrm{d}z\mathrm{d}
v(Aa_g)_{i}G_{i}(I-P)g_{i}-\int_{\mathbb{R}\times \mathbb{R}^{3}}
\hskip-5pt\mathrm{d}z\mathrm{d}v\frac{1}{Mw_{i}}(I-P)g_{i}G_{i}(I-P)g_{i} \\
&&+\int_{\mathbb{R}\times \mathbb{R}^{3}}\hskip-5pt\mathrm{d}z\mathrm{d}v\frac{1}{Mw_{i}}(I-P)g_{i}L(I-P)g_{i}\Bigg\}+\langle \Gamma,Aa_g\rangle
+\la \frac{1}{Mw}(I-P)g,\Gamma\ra.
\end{eqnarray*}
The first term on the right hand side vanishes since $w_i(Aa_g)_{i}\partial
_{z}(Aa_g)_{i}$ are functions of $z,t$ only and the Maxwellian is centered. By recalling the definition of $
G_{i}$, (\ref{GG}),
\begin{equation*}
G_{i}(I-P)g_{i}=v_{z}\partial _{z}(I-P)g_{i}+U\ast w_{j}^{\prime }\partial
_{v_{z}}(I-P)g_{i},
\end{equation*}
we have for the third term
\begin{eqnarray*}
-\int_{\mathbb{R}\times \mathbb{R}^{3}}\mathrm{d}z\mathrm{d}
v(Aa_g)_{i}G_{i}(I-P)g_{i} &=&-\int_{\mathbb{R}\times \mathbb{R}^{3}}\hskip-5pt\mathrm{d}z
\mathrm{d}v(Aa_g)_{i}v_{z}\partial _{z}(I-P)g_{i} \\
&=&\int_{\mathbb{R}\times \mathbb{R}^{3}}\mathrm{d}z\mathrm{d}v\partial
_{z}(Aa_g)_{i}v_{z}(I-P)g_{i}
\end{eqnarray*}
which cancels with the second term ($-\int_{\mathbb{R}\times \mathbb{
R}^{3}}\mathrm{d}z\mathrm{d}v(I-P)g_{i}v_{z}\partial _{z}(Aa_g)_{i}$) in the
right hand side. By using the definition of $G_{i}$ we get for the fourth
term
\begin{eqnarray*}
&&-\int_{\mathbb{R}\times \mathbb{R}^{3}}\mathrm{d}z\mathrm{d}v\frac{1}{
Mw_{i}}(I-P)g_{i}G_{i}(I-P)g_{i} \\
&=&-\int_{\mathbb{R}\times \mathbb{R}^{3}}\mathrm{d}z\mathrm{d}v\frac{1}{
Mw_{i}}\frac{1}{2}\left[ v_{z}\partial _{z}((I-P)g_{i})^{2}+U\ast
w_{j}^{\prime }\partial _{v_{z}}((I-P)g_{i})^{2}\right] \\
&=&-\int_{\mathbb{R}\times \mathbb{R}^{3}}\mathrm{d}z\mathrm{d}v\frac{v_{z}}{
2Mw_{i}}[\frac{w_{i}^{\prime }}{w_{i}}+\beta U\ast w_{j}^{\prime
}]((I-P)g_{i})^{2}=0
\end{eqnarray*}
by using the equation for the front. \qed

\bigskip
In the next Lemmas we apply above identity and the estimates in Section 3 to bound the weighted norms of $h$ and its space and time derivatives.
\bigskip
\begin{lm}
\label{0weight} Let $0\le\gamma$ be sufficiently small. Then if $\|h(t)\|_{M,\gamma }\leq \delta _{0}$
\begin{equation}
\frac12\frac{d}{dt}\|h(t)\|_{M,\gamma }^{2}+\nu_0 \|h(t)\|_{D,\gamma }^{2}\leq
C(\gamma +\delta _{0})(\|\partial _{t}h(t)\|_{D,\gamma
}^{2}+\|h(t)\|_{D,\gamma }^{2}),
\label{0weightf}
\end{equation}
with $\nu_0$  given in Lemma \ref{lgap}.
\end{lm}

\noindent \textbf{Proof.} Note that $g=\zg h$ satisfies%
\begin{equation}\label{defgammai}
\partial _{t}g_{i}+G_{i}g_{i}-Lg_{i}=F_{i}(h)\partial _{v_{z}}g_{i}+\hat{G} h_{i}+\frac{2zv_z\gamma g_{i}}{1+z^{2}}\equiv \Gamma _{i}.
\end{equation}
where
\begin{equation}
\hat{G}h_{i}=v_{z}Mw_{i}\beta \int d{\rm z}^{\prime }  U^{\prime }(z-z^{\prime
})(\zg-\zgp ) (a_h)_{j}(\,z^{\prime
}, t)\ .  \label{hatg}
\end{equation}
We now apply Lemma \ref{identity}. We first treat $F(a_h)\partial
_{v_{z}}g$. Notice that \[\la F(a_h)\partial _{v_{z}}g,(Aa_g)\ra=0,\]
and
\begin{eqnarray*}
&&\left|\sum_{i=1}^{2}(F_{i}(h)\partial _{v_{z}}g_{i},\frac{1}{Mw_{i}}(I-P)g_{i})\right|
\\
&\leq &C\Big[\|F(h)\|_{L_{\infty }}\ \big( \|a_{g}\| +
\|\partial_{v_z}(I-P)g\|_{M}\big)\Big]\|(I-P)g\|_{M} \\
&\leq &C\big[\|\partial _{t}h\|_{D}+\|h\|_{D}\big]\|g\|_M\  \|g\|_{D}+\|h\|_M\
\|g\|_{D}^{2} \\
&\leq &C\delta _{0}\big[\|h\|_{D}+\|\partial _{t}h\|_{D}\big]^{2}+C\delta
_{0}\|g\|_{D}^{2}.
\end{eqnarray*}
Next we estimate $\hat{G}h_{i}$. Note that $\langle \hat{G}h,Aa_{g}\rangle =0$.
By (\ref{z-z'}),  
$$|\zg-\zgp|\leq C\gamma
|z-z^{\prime }|(z_{\g-\frac 1 2}+z'_{\g-\frac 1 2})\le C\gamma
|z-z^{\prime }|$$
for $\gamma<\frac{1}{2} $, recalling $
a_{h}=a_{h}^{(1)}+a_{h}^{(2)}$, we deduce that
\begin{eqnarray*}
&&(\hat{G}h_i,\frac{1}{Mw_{i}}(I-P)g_{i})  \\
&=&\beta \int dv \ v_{z}(I-P)g_i\left(\int d{\rm z}'\ U\partial_z(z-z^{\prime })[\zg-\zgp](a_{h})_j(\,z^{\prime }, t)\right) \\
&= &\beta \int dv \ v_{z}(I-P)g_i\left(\int d{\rm z}'\ \partial_z U(z-z^{\prime
})[\zg -\zgp](a_{h}^{(2)})_j(\,z^{\prime }, t)]\right) \\
&&+\beta \int dv\  v_{z}(I-P)g_i\left(\int d{\rm z}'\ U(z-z^{\prime })\partial _{z^{\prime
}}\big[(\zg -\zgp)(a_{h}^{(1)})_j(\,z^{\prime }, t)\big]\right)\ ,
\end{eqnarray*}
Hence
\begin{eqnarray*}
&&\left|(\hat{G}h_i,\frac{1}{Mw_{i}}(I-P)g_{i})\right| \\
&\leq &C\gamma \Big(\|a_{h}^{(2)}\|+\|\partial _{z}a_{h}^{(1)}\|+\|U*
\frac{za_{h}^{(1)}}{(1+z^{2})^{1-\gamma }}\|\Big)\|(I-P)g\|_M \\
&\leq &C\gamma [\|g\|_{D}^{2}+\|\partial_t g\|_{D}^{2}]\ .
\end{eqnarray*}
 We have used (\ref{agamma}) and \ref{a2gamma} in Lemma \ref{0hydro}.

 For the third term $\displaystyle{\frac{2zv_z\gamma g_{i}}{1+z^{2}}}$ in the definition of $\G_i$, (\ref{defgammai}), since $v_zM^{-1}$ $=\beta^{-1}\partial_{v_z} M^{-1}$, an integration by part in the  $v-$variable and  again
estimate (\ref{agamma}) give
\begin{eqnarray*}
&&\sum_{i=1}^{2}\left(\frac{2zv_z\gamma g_{i}}{1+z^{2}},\frac{1}{Mw_{i}}
(I-P)g_{i}\right) \\
&=&\sum_{i=1}^{2}\left(\frac{2zv_z\gamma (I-P)g_{i}}{1+z^{2}},\frac{1}{Mw_{i}}
(I-P)g_{i}\right)\\&+&\sum_{i=1}^{2}\left(\frac{2zv_z\gamma Pg_{i}}{1+z^{2}},\frac{1}{  Mw_{i}
}(I-P)g_{i}\right)\end{eqnarray*}
Hence
\begin{eqnarray*}
&&\left|\sum_{i=1}^{2}\left(\frac{2zv_z\gamma g_{i}}{1+z^{2}},\frac{1}{Mw_{i}}
(I-P)g_{i}\right)\right|\\
&\leq & C\gamma\sum_{i=1}^{2}\left(\frac{2z (I-P)g_{i}}{1+z^{2}},\frac{1}{\beta Mw_{i}}
\partial_{v_z}(I-P)g_{i}\right)+C\gamma \big[\|\partial _{t}g\|_{D}^{2}+\|g\|_{D}^{2}\big]\\
&\leq & C\gamma \big[\|\partial _{t}g\|_{D}^{2}+\|g\|_{D}^{2}\big]\ .
\end{eqnarray*}

\noindent On the other hand,
\begin{eqnarray*}
\left|\left\langle \frac{2zv_z\gamma (I-P)g}{1+z^{2}}, Aa_{g}\right\rangle  \right|
\leq C\gamma\|(I-P)g\|_{M}\|\frac{z}{1+z^2}Aa_{g})\|.
\end{eqnarray*}
We use the splitting
$$\left\|\frac{z}{1+z^2}Aa_{g})\right\|\le 
\left\|\frac{z}{1+z^2}Aa^{(2)}_{g})\right\|+\left\|\frac{z}{1+z^2}Aa^{(1)}_{g})\right\|\ $$
and introduce, as usual, the commutator to get
$$\left\|\frac{z}{1+z^2}Aa^{(2)}_{g})\right\|\le \left\|\frac{z\zg}{1+z^2}Aa^{(2)}_{})\right\|+ \left\|\frac{z}{1+z^2}[U,\zg]a^{(2)}_{})\right\|\ .$$
The above term is immediately bounded by using he definition of $Aa^{(2)}_h$, the second by using Theorem \ref{hydro}.

\noindent As for the contribution from $a^{(1)}_h$, we have 
\begin{eqnarray*}
\left\|\frac{z}{1+z^2}Aa^{(1)}_{g})\right \|\  &\le& \left\| A\left(\frac{z\zg}{1+z^2}a^{(1)}_{h}\right)\right\|+\left\| [A, \frac{z}{1+z^2}]\zg a^{(1)}_{h}\right\| \\ 
&\le &\displaystyle{ C(\|\partial _{t}h\|_{D,\gamma } +\|h\|_{D,\gamma })}\ .
\end{eqnarray*}
Inded, the first term is bounded by using the boundedness of $A$ and (\ref{agamma}). To bound the commutator, we use
\begin{equation}\label{commut}
\left[\frac{z}{1+z^2},A\right]\zg a^{(1)}_{h}=\int_\R d{\rm z}'\  U(z-z')\left(\frac{z}{1+z^2}-\frac{z'}{1+z'^2}\right)\zgp a^{(1)}_h(z'),
\end{equation}
the inequality
\begin{equation}\label{z-z'1}
\left|\frac{z}{1+z^2}-\frac{z'}{1+z'^2}\right|\le C\frac{|z-z'|}{\zhal\zphal},
\end{equation}
and (\ref{pointwise}).

Therefore 
$$\displaystyle{\left|\left\langle \frac{2zv\gamma (I-P)g}{1+z^{2}},Aa_{g}\right\rangle\right| }\le 
C\gamma \|g\|_{D} (\|\partial _{t}h\|_{D,\gamma }+\|h\|_{D,\gamma })
$$
and this concludes the proof of the lemma.\qed

\vskip.1cm

We notice that in the proof of this Lemma we are allowed to apply   (\ref{agamma}) since we are assuming $\gamma$ small enough and hence also $\gamma\le\frac{1}{8}$.

\bigskip

\begin{lm}
\label{tweight} If $\|\partial h(t)\|_{M,\gamma }+\|h(t)\|_{M }\leq
\delta _{0}$, then
\begin{equation}
\frac12\frac{d}{dt}\|\partial _{t}h(t)\|_{M,\gamma }^{2}+\nu_0\|\partial
_{t}h(t)\|_{D,\gamma }^{2}\leq C(\gamma +\delta _{0})\|\partial
h(t)\|_{D,\gamma}^{2}+C\|h(t)\|_{D,\gamma -\frac{1}{2}
}^{2}.
\label{tweightf}
\end{equation}
\end{lm}
\bigskip
\noindent\textbf{Proof.} Let $g=\zg \partial _{t}h$. We have
\begin{equation}\label{defgammai1}
\lbrack \partial _{t}+G_{i}-L]g_{i}=\frac{2zv_z\gamma g_{i}}{1+z^{2}}+
\hat{G}\partial _{t}h_{i}+F_{i}(h)\partial _{v_{z}}g_{i}+F_{i}(\partial _{t}h)\partial _{v_{z}}h_{i}\zg \equiv\Gamma_i.
\end{equation}
By Lemma \ref{identity} we need to estimate
$$\langle Aa_{g},\Gamma \rangle+\langle
\frac{1}{Mw}(I-P)g,\Gamma\rangle$$
We first estimate the contribution due to $\displaystyle{\frac{2\gamma zv_zg}{1+z^{2}}}$. By using again $g=a_{g}M+(I-P)g$,
\begin{equation*}
\left\langle\frac{2\gamma zv_zg}{1+z^{2}},Aa_{g}\right\rangle =\left\langle \frac{2\gamma
zv_za_{g}M}{1+z^{2}},Aa_{g}\right\rangle +\left\langle \frac{2\gamma zv_z(I-P)g}{1+z^{2}},Aa_{g}\right\rangle.
\end{equation*}
By the same argument used in Lemma \ref{0weight}  we have
 \begin{eqnarray*}
&&\left\|\frac{ z}{1+z^2}Aa_{g}\right\|\leq \left\|A\left(\frac{z a_{g}}{1+z^2}\right)\right\|\hskip-.1cm+\hskip-.1cmC\hskip-.1cm\left\|\int \frac{
U(z-z')|z-z'|}{\zhal\zphal}a_{g}(z')dz'\right\| \\
&\leq &C\|(1+z^2)^{\gamma -\frac{1}{2}}\partial _{t}a_{h}\|\leq
C\|(I-P)\partial _{z}h\|_{M,\gamma -\frac{1}{2}}.
\end{eqnarray*}
The last inequality is due to (\ref{a}).
 We
therefore have
\begin{equation*}
\left|\left\langle\frac{2\gamma zv_zg}{1+z^{2}},Aa_{g}\right\rangle \right|\leq \varepsilon
\|\partial_t h\|_{D,\gamma }^{2}+C_{\varepsilon }\gamma \|\partial_z h\|_{D,\gamma -\frac{1}{2}
}^{2}.
\end{equation*}
We now estimate $\displaystyle{|\la\frac{1}{Mw}(I-P)g,\frac{zv_z\gamma g}{
1+z^{2}}\ra|}$. As before, an integration by part in the  $v$-variable provides
\begin{eqnarray}
&&\left|\left\la \frac{1}{Mw}(I-P)g,\frac{zv_z\gamma g}{
1+z^{2}}\right\ra\right| \nonumber\\
&\leq &\left|\left\la\frac{1}{Mw}(I-P)g,\frac{zv_z\gamma
(I-P)g}{1+z^{2}}\right\ra\right| \nonumber \\&+& \left|\left\la \frac{1}{Mw}(I-P)g,\frac{zv_z\gamma
(1+z^{2})^{\gamma }P\partial _{t}h}{1+z^{2}}\right\ra\right|   \\
&\leq &C\gamma \left|\left\la\frac{1}{\beta\zhal}(I-P)g, \partial_{v_z}(I-P)g\right\ra_{M}\right |\nonumber
\label{1-pg}\\&
+&
C\gamma \|(I-P)g\|_{M}^{2}+C_{\varepsilon }\|(1+z^{2})^{\gamma
-\frac 1 2}\partial _{t}a_{h}\|^{2} \nonumber\\
&\leq &C\gamma \|\partial h\|_{D,\gamma -\frac{1}{2}}^{2}.\nonumber
\end{eqnarray}

\noindent We estimate  the second term $\hat{G}\partial_{t}h_{i}$ in $\Gamma$ by first noting that 
$\la\hat{G}\partial_{t}h,(Aa_{g})\ra=0$. With an argument similar to the one used to estimate $\hat Gh$, we obtain
\begin{eqnarray*}
\left|\left\la\frac{1}{Mw}(I-P)g,\hat{G}\partial _{t}h\right\ra\right |
\leq \varepsilon \|g\|_{D}^{2}+C_{\varepsilon }\gamma \|\partial
_{t}h\|_{D,\gamma -\frac{1}{2}}^{2}.
\end{eqnarray*}

As for the third term in $\G$, $F(h)\partial _{v_{z}}g$, we note that \[
\la F(h)\partial _{v_{z}}g,(Aa_{g})\ra =0,\] and
\begin{eqnarray*}
&&\left\la F(h)\partial _{v_{z}}g,\frac{1}{Mw}(I-P)g\right\ra
\\
&=& \left\la F(h)\partial _{v_{z}}Pg,\frac{1}{Mw}
(I-P)g\right\ra
+ \left\la F(h)\partial _{v_{z}}(I-P)g,\frac{1}{
Mw}(I-P)g\right\ra \\
&\leq &C\|F_{i}(h)\|_{\infty }\left(\|a_{g}\|\cdot \|(I-P)g_{i}\|_{M}+\|\partial_{v_z}
(I-P)g\|_{M}^{2}\right)\ .\end{eqnarray*}
Hence
\begin{eqnarray*}
\left|\left\la F(h)\partial _{v_{z}}g,\frac{1}{Mw}(I-P)g\right\ra\right|
\leq C\|\partial _{z}a_{h}\|\left(\|a_{g}\|\cdot \|g\|_{D}+\|g\|_{D}^{2}\right).
\end{eqnarray*}

To estimate the fourth term $z_{\gamma }F(\partial
_{t}a_{h})\partial _{v_{z}}h$, we first remind that
\begin{equation*}
\la \zg F(\partial _{t}a_{h})\partial
_{v_{z}}h,Aa_{g}\ra=0.
\end{equation*}

Since by (\ref{a}) 
\[
\|\zg F(\zg \partial _{t}a_{h})\|_\infty
\leq C\|(I-P)\partial_{z}h\|_{M ,\gamma}\ ,
\]
and
\[\|[\zg, F] \partial _{t}a_{h})\|_\infty
\leq C\|(I-P)\partial_{z}h\|_{M ,\gamma-\frac{1}{2}}\ ,\]
we have, by using the smallness assumption and integrating by part on $v$,
\begin{eqnarray*}
\left|\left\la z_{\gamma }F(\partial
_{t}a_{h})\partial _{v_{z}}h,\frac{1}{Mw}(I-P)g\right\ra\right|&\leq&
C\|h\|_{M}\ \|\zg F(\partial_t a_h)\|_\infty\  \|\partial
_{t}h\|_{D }\\&\le& C\delta _{0}\|\partial h\|^2_{D,\gamma}.
\end{eqnarray*}
This concludes the proof of the Lemma. \qed
\vskip.2cm
\begin{lm}\label{zweight}
If $\|\partial  h(t)\|_{M }+\|h(t)\|_{M}\leq \delta _{0}$, then
\begin{eqnarray}
&&\frac12\frac{d}{dt}\|\partial _{z}h(t)\|_{M}^{2}+\nu_0 \|\partial
_{z}h(t)\|_{D }^{2} \leq  C\left(\|h\|_{D}^{2}+\|\partial_t  h\|_{D}^{2}
+\delta_0\|\partial h\|^2_{D}\right)\ .
\label{zweightf} 
\end{eqnarray}
Moreover, given  $0< \gamma \leq 1$,  if $
\|\partial  h(t)\|_{M,\gamma }+\|h(t)\|_{M}\leq \delta _{0}$ and $\|\partial_z h\|_{M,\frac 1 2}<\eta$, then
\begin{eqnarray}
&&\frac12\frac{d}{dt}\|\partial _{z}h(t)\|_{M,\gamma }^{2}+\nu_0 \|\partial
_{z}h(t)\|_{D,\gamma }^{2}\nonumber\\
\qquad &&\leq  C\left(\|h\|_{D,\gamma -\frac{1}{2}
}^{2}+\|\partial_t  h\|_{D,\gamma -\frac{1}{2}}^{2}
+\delta_0\|\partial h\|^2_{D,\gamma}+\gamma \|\partial_z  h\|_{D,\gamma -\frac{1}{2}}^{2}\right)\ .
\label{zweightf1}
\end{eqnarray}
\end{lm}

\bigskip
\noindent \textbf{Proof.} We define $g=\zg \partial _{z}h$ to get
\begin{eqnarray*}
&&\partial _{t}g_{i}+G_{i}g_{i}-Lg_{i} =\frac{2\gamma zv_zg_{i}}{1+z^{2}}+\zg \beta v_z Mw_i'U\ast\partial
_{z}a_{j} \\
&&+\hat{G}\partial _{z}h_{i}+\partial _{z}U\ast w_{j}^{\prime }\partial
_{v_{z}}(\zg h_{i})-F_{i}(h)\partial
_{v_{z}}g_{i}-\zg F_{i}(\partial _{z}a_{h})\partial
_{v_{z}}h_{i}
\equiv \Gamma _{i}.
\end{eqnarray*}
where $\hat{G}$ is defined in (\ref{hatg}). Since
$g=a_{g}M+(I-P)g$, by  Lemma \ref{identity} we need
to estimate 
$$\langle \Gamma ,Aa_{g}\rangle +\displaystyle{\left\la\frac{1}{
Mw}(I-P)g,\Gamma \right\ra }.$$
In this proof, for consistency with the notation in Lemma \ref{azz}, we switch from  $a_g$ to $a_\g$.
We first estimate the first term $\displaystyle{\frac{2\gamma zv_zg_{i}}{1+z^{2}}}$.
\begin{equation*}
\left\langle \frac{2\gamma zv_zg}{1+z^{2}},Aa_{\g}\right\rangle =\left\langle \frac{2\gamma
zv_za_{\g}M}{1+z^{2}},Aa_{\g}\right\rangle +\left\langle \frac{2\gamma zv_z(I-P)g}{1+z^{2}}
,Aa_{\g}\right\rangle .
\end{equation*}
The first contribution above vanishes. For the second term, we split
 $a_{\g}=a_{\g}^{(1)}+a_{g}^{(2)}$. Then we have
 \begin{equation*}
\left \|\frac{z}{1+z^2}Aa^{(2)}_{\g}\right\|\le C\|\partial_zh\|_{D, \gamma-\frac{1}{2}}
\end{equation*}
by an argument  similar to the one used in (\ref{zgammadza2h}) --  (\ref{zgammadza2ht}).
On the other hand,
\begin{equation*}
\frac{z}{1+z^2}Aa^{(1)}_{\g}=A\left(a^{(1)}_{\g}\frac{z}{1+z^2}\right)+\left[\frac{z}{1+z^2},A\right]a^{(1)}_{\g}.
\end{equation*}
By the boundedness of $A$,
\[
\left\|A(a^{(1)}_{\g}\frac{z}{1+z^2})\right\|\le C \|\zga  \partial_z a_{h}^{(1)}\|.
\]
The commutator can be estimated in the usual way and we conclude that
\[
\left\|[\frac{z}{1+z^2},A]a^{(1)}_{\g}\right\|\le C\|\zga\partial_za^{(1)}_{h}\|.
\]
Collecting all the estimates and using Lemma \ref{0hydro} to bound $\|\zga\partial_za^{(1)}_{h}\|$, we have
\[
\left\|\frac{z}{1+z^2}Aa^{(1)}_{\g}\right\| \le C\left(\|h\|_{D,\gamma -\frac{1}{2}}+\|\partial h\|_{D,\gamma -\frac{1}{2}}\right).
\]
Therefore, for any $\e>0$ there is $C_\e$ such that
\[
\left |
\left\langle \frac{2\gamma zv_z(1-P)g}{1+z^{2}},Aa_{\g}\right\rangle\right|\le \e \|\partial_z h\|_{D,\g}^2+C_\e\gamma(\|h|_{D,\gamma-\frac{1}{2}}^2+\|\partial h\|_{D,\g-\frac{1}{2}}^2).
\]
We deal with next term as we already did to get (\ref{1-pg}): Using $v_zM^{-1}=\beta^{-1}\partial_{v_z} M^{-1}$, an integration by part in the \thinspace $v-$variable, provides\begin{eqnarray*}
&&\left|\left\la\frac{1}{Mw}(I-P)g,\frac{zv_z\gamma g}{1+z^{2}}
\right\ra\right| \\
&\leq &\left|\left\la\frac{1}{Mw}(I-P)g,\frac{zv_z\gamma
(I-P)g}{1+z^{2}}\right\ra\right| +\left|\left\la\frac{1}{Mw}(I-P)g,\frac{zv_z\gamma
\zg P\partial _{z}h}{1+z^{2}}\right\ra\right| \\
&\leq &C\gamma \|\frac{1}{1+|z|}(I-P)g\partial_{v_z}(I-P)g\|_{M}+\varepsilon \|(I-P)g\|_{M}^{2}+C_{\varepsilon }\gamma
\|\zga \partial _{z}a_{h}\|^{2} \\
&\leq &\varepsilon \|(I-P)g\|_{M}^{2} +C_{\varepsilon }\gamma \left(\|h\|_{D,\gamma -\frac{1}{2}}^{2}+\|\partial
h\|_{D,\gamma -\frac{1}{2}}^{2}+\|h\|_{D}^{2}+\|\partial
h\|_{D}^{2}\right).
\end{eqnarray*}
In the last inequality we have used Lemma \ref{0hydro} and Lemma \ref{azz}.

The second term is easily obtained  by Theorem \ref{hydro}. Indeed, 
$$\langle \zg \beta v_z Mw_i'U\ast\partial
_{z}a_{j} ,(Aa_{\g})_i\rangle=0$$ 
and 
\begin{eqnarray*}&&\left|\langle \zg \beta v_z  \frac{w'_i}{
w}U\ast\partial
_{z}a_{j},(I-P)g_i\rangle\right|\\&&
\le \varepsilon \|(I-P)g\|_{M}^{2}+C_{\varepsilon }\left(\|\partial
_{z}a^{(1)}_{h}\|^{2}+\|a^{(2)}_{h}\|^{2}\right)\\&&
\le \varepsilon \|(I-P)g\|_{M}^{2}+C_{\varepsilon }\left(\|\partial_t h\|^2_{D}+\|h\|_{D}^{2}\right)\ .
\end{eqnarray*}
\noindent We now estimate the fourth term $\zg \partial _{z}U\ast
w_{j}^{\prime }\partial _{v_{z}}h_{i}$. Notice that

\begin{equation*}
\la \zg \partial _{z}U\ast w^{\prime }\partial
_{v_{z}}Ph,(Aa_{\g})\ra =0\ .
\end{equation*}
Introducing as usual the commutator, we obtain
\begin{eqnarray*}
\left|\left\la \zg \partial _{z}U\ast w^{\prime }\partial
_{v_{z}}(1-P)h,Aa_{\g}\right\ra\right| &\le& \left|\left\la \zg \partial _{z}U\ast w^{\prime }\partial
_{v_{z}}(1-P)h,\zg A\partial_z a_{h}\right\ra\right|\\&+&\left|\left\la \zg \partial _{z}U\ast w^{\prime }\partial
_{v_{z}}(1-P)h,[A,\zg]\partial_za_{h}\right\ra\right|\ .
\end{eqnarray*}
Then we have  the bound $|\zg^2w'|\le C$ because of the fast decay of $w'$ 
and   $\|A\partial_z a_h\|\le C(\|h\|_D+\|\partial h\|_D)$ by using
(\ref{ag2}) and (\ref{azgamma}). The commutator term is bounded by using (\ref{stimacomm}) to obtain $\|[A,\zg]\partial_z a_h\|\le C\|\partial_z a_h\|_{\g-\frac12}$ which is then bounded again by using (\ref{ag2}) and (\ref{azgamma}). In conclusion we obtain:
\begin{eqnarray*}
\left|\left\la \zg \partial _{z}U\ast w^{\prime }\partial
_{v_{z}}(1-P)h,Aa_{\g}\right\ra\right|&\leq &C\left(\|h\|_{D,\gamma -\frac{1}{2}
}+\|\partial h\|_{D,\gamma -\frac{1}{2}}\right)\|h\|_D.
\end{eqnarray*}
On the other hand,
\begin{eqnarray*}
&&\la\zg \partial _{z}U\ast w^{\prime
}\partial _{v_{z}}h ,\frac{1}{Mw }(I-P)g \ra \\
&=&\la\zg \partial _{z}U\ast w^{\prime
}\partial _{v_{z}}(I-P)h ,\frac{1}{Mw }(I-P)g \ra\\&+&\la\zg \partial _{z}U\ast w^{\prime
}\partial _{v_{z}}Ph ,\frac{1}{Mw }(I-P)g \ra.
\end{eqnarray*}
The first term is clearly bounded by $\varepsilon
\|(I-P)g\|_{M}^{2}+C_{\varepsilon }\|h\|_{D}^{2}$.
For the second
term,
\begin{eqnarray*}
&&\left|\sum_{i=1}^{2}(\zg \partial _{z}U\ast w_{j}^{\prime
}\partial _{v_{z}}Ph,\frac{1}{Mw}(I-P)g)\right| \\
&\leq &\varepsilon \|(I-P)g\|_{M}^{2}+C_{\varepsilon }\left(\|\partial_t h\|^2_{D}+\|h\|_{D}^{2} +\|a^{(2)}_{h}\|^{2}+\|h\|_{D}^{2}\right).
\end{eqnarray*}
To get the last inequality we have decomposed, as usual $a_h=a_h^{(1)}+ a_h^{(2)}$ and bounded the contribution due to $a_h^{(2)}$ by the first part of Theorem \ref{hydro} and the one due to $a_h^{(1)}$ by using (\ref{pointwise}) and the fast decay of $w'$.
Then, by using Theorem \ref{hydro} we get the final bound
\begin{eqnarray*}
\varepsilon \|(I-P)g\|_{M}^{2}&+&C_{\varepsilon }\left(\|\partial
_{z}a^{(1)}_{h}\|^{2}+\|a^{(2)}_{h}\|^{2}+\|h\|_{D}^{2}\right)\\
&\le &
\varepsilon \|(I-P)g\|_{M}^{2}+C_{\varepsilon }\left(\|\partial_t h\|^2_{D}+\|h\|_{D}^{2}\right)\ .
\end{eqnarray*}
\noindent Now turn to the third term $\hat{G}h$ in $\Gamma $. Since $\langle \hat{G}h,Aa_{\g}\rangle =0$,
\begin{eqnarray*}
&&\left|\left\la\frac{1}{Mw}(I-P)g,\hat{G}\partial
_{z}a_{h}\right\ra\right| \\
&\leq &\varepsilon \|(I-P)g\|_{M}^{2}+C_{\varepsilon }\|\partial
_{t}h\|_{D,\gamma -\frac{1}{2}}^{2}+C_{\varepsilon }\|h\|_{D,\gamma -%
\frac{1}{2}}^{2}.
\end{eqnarray*}

\noindent For the fifth term $F(a_h)\partial _{v_{z}}g$, we note that
$\la F(a_h)\partial _{v_{z}}g,Aa_{\g}\ra =0$ and
\begin{eqnarray*}
&&\hskip-.1cm\left|\hskip-.1cm\left\la F(a_h)\partial _{v_{z}}g,\frac{1}{Mw}
(I-P)g\right\ra \hskip-.1cm\right| \\
&\le& \hskip-.15cm\left|\hskip-.1cm\left\la F(a_h)\partial _{v_{z}}Pg,\hskip-.1cm\frac{1}{Mw}
(I-P)g) \right\ra \hskip-.1cm\right| \hskip-.1cm+\hskip-.1cm\left|\hskip-.1cm\left\la F(a_h)\partial
_{v_{z}}(I-P)g,\hskip-.1cm\frac{1}{Mw}(I-P)g\right\ra \hskip-.1cm\right|  \\
&\leq &C\|F(a_h)\|_{\infty }\left(\|a_{\g}\| \|(I-P)g\|_{M}+\|\partial_{v_z}(I-P)g\|_{M}^{2}\right) \\
&\leq &C\|\partial _{z}a_{h}\|\ \left(\|a_{\g}\|\|(I-P)g_{i}\|_{M}+\|\partial_{v_z}(I-P)g\|_{M}^{2}\right) \\
&\leq &C\delta _{0}\left(\|g\|_{D}^{2}+\|h\|_{D}^{2}+\|\partial
_{t}h\|_{D}^{2}\right).
\end{eqnarray*}

Finally, to estimate  the sixth term $\zg F(\partial
_{z}a_{h})\partial _{v_{z}}h$, we note
\begin{equation*}
\int_{\R^3} dv\ \zg F_{i}(\partial _{z}a_{h})\partial
_{v_{z}}h_{i}(Aa_{\g})_i=0.
\end{equation*}
To treat the last term we consider separately the case $\gamma=0$ and the case $\gamma>0$.
In the first case we simply get, 
\begin{eqnarray*}
&&\left|\left\la F(\partial
_{z}a_{h})\partial_{v_{z}}h,\frac{1}{Mw}(I-P)\partial_z h\right\ra \right| \\&\leq &C\left(\| \partial_z a_{h}\| \|a_{h}\|+\|\partial_z a_{h}\|
\|h\|_{D}\right)\|(I-P)\partial_z h\|_{M}\\ &\leq &C\delta_{0}\|\partial h\|_{D}^{2}
\end{eqnarray*}
by using (\ref{azgamma}) and (\ref{ag2}) with $\g=0$ to bound $\|\partial_z a_h\|$.

In the case $\g>0$ we need to employ Lemma \ref{azz} to treat the last term as
\begin{eqnarray*}
&&\left|\left\la \zg F(\partial
_{z}a_{h})\partial _{v_{z}}h,\frac{1}{Mw}(I-P)g\right\ra \right| \\
&\le&\left|\left\la F(a_{\g})\partial _{v_{z}}h,\frac{1}{Mw}
(I-P)g\right\ra\right|
+\left|\left\la( \lbrack \zg ,F](\partial
_{z}a_{h})\partial _{v_{z}}h,\frac{1}{Mw}(I-P)g\right\ra\right| \\
&\leq &\left(\|a_{\g}^{(2)}\|\cdot \|a_{h}\|+\|a_{\g}\|\cdot
\|h\|_{D}\right)\|(I-P)g\|_{M} \\
&&+\left((\|\partial _{z}\{a_{\g}^{(1)}-\zg a_{h}^{(3)}\}\|+\|\zg a_{h}^{(3)}\|)\cdot
\|a_{h}\|\right)\|(I-P)g\|_{M} \\
&&+\left(\|\zga \partial _{z}a_{h}^{(1)}\|\cdot
\|a_{h}\|\right)\|(I-P)g\|_{M} \\
&\leq &C\delta _{0}\Big(\|g\|_{D}^{2}+\|h\|_{D}^{2}+\|\partial
 h\|_{D,\gamma}^{2}+\|h\|_{D,\gamma -\frac{1}{2}}^{2}+\|\partial
_{t}h\|_{D,\gamma -\frac{1}{2}}^{2}\Big).
\end{eqnarray*}
We deduce our lemma by letting $\varepsilon $ small and using $\delta_0$ small. \qed

We remark that this is the  only point where we use (\ref{ag1}). The relevance of this estimate is in the fact that we get a bound involving the norm of the function with a power  $\gamma-\frac{1}{2}$. This is crucial for the final consistency argument.
\vskip.2cm
\noindent\textbf{Proof of Theorem }\ref{result}: To prove the first part, we start
with $\gamma =0$ in all three Lemmas \ref{0weight}, \ref{tweight} and \ref{zweight}. We multiply by a positive number $K$ (\ref{0weightf}) and (\ref{tweightf}) and add both to (\ref{zweightf}):

\begin{eqnarray*}&&\frac12\frac{d}{dt}\Big(K\big(\|h(t)\|_{M }^{2}+\|\partial _{t}h(t)\|_{M }^{2}\big)+\|\partial_z h(t)\|^2_M\Big)\\ &&+K\nu_0\Big( \|h(t)\|_{D }^{2}+\|\partial_{t}h(t)\|_{D}^{2}\Big)+\nu_0\|\partial _{z}h(t)\|_{D}^{2}
\\ &&\leq
C\Big(K\delta_0\left(\|h(t)\|_{D
}^{2}+\|\partial
 h(t)\|_{D}^{2}\right)+\|h(t)\|_{D}^{2}\\ &&+\|\partial
_{t}h(t)\|_{D}^{2}+\delta_0\|\partial
_{z}h(t)\|_{D}^{2}\Big)\end{eqnarray*}

\noindent By choosing  $K>\displaystyle{\frac{C}{4\nu_0}}$, and $\delta_0<\displaystyle{\frac{\nu_0}{4C}}$, we obtain that
\begin{eqnarray}
&&\frac12\frac{d}{dt}\Big(K\left(\|\partial _{t}h\|^{2}_M+\|h\|^{2}_M\right)+\|\partial _{z}h\|^{2}_M\Big)\nonumber\\&&+\frac{\nu_0 }{2}\Big(K\left(\|\partial
_{t}h\|_{D}^{2}+\|h\|_{D}^{2}\right)+\|\partial
_{z}h\|_{D}^{2}\Big)\leq 0.  \label{largek}
\end{eqnarray}
Then, a standard continuity argument shows that  the assumption 
$$\|h(t)\|_{M}^2+\|\partial h(t)\|_{M}^2\le\delta_0$$
is verified at any time $t$, thus 
 completing  the proof of the first part of Theorem \ref{result}. In particular, we have
\begin{equation}
\int_0^\infty dt \Big(K\left(\|\partial_t
h(t)\|_{D}^{2}+\|h(t)\|_{D}^{2}\right)+\|\partial_z
h(t)\|_{D}^{2}\Big)\le C(\|h(0)\|_{M}^2+\|\partial h(0)\|_{M}^2)
\label{ult?}
\end{equation}

To prove the second part, we first prove an inequality  like (\ref{largek}) for the weighted norms with weight $z_{\gamma_0}$, for $\gamma_0$ small.  We will use a standard continuity argument with the assumption
\begin{equation}
(\|h(t)\|_{M,\gamma_0}^2+\|\partial h(t)\|_{M,\frac12+\gamma_0}^2)\le\delta_0.
\label{gammasmall}
\end{equation}
As first step,  we multiply once again (\ref{0weightf}) and (\ref{tweightf}) by $K$ and add them to (\ref{zweightf1})
\begin{eqnarray*}
\frac12\frac{d}{dt}&\Big(&K(\|h(t)\|_{M,\gamma_0 }^{2}+\|\partial _{t}h(t)\|_{M,\gamma_0 }^{2})+\|\partial _{z}h(t)\|_{M,\gamma_0 }^{2}\Big)\\
&+& K\nu_0\Big( \|h(t)\|_{D,\gamma_0 }^{2}+\partial
_{t}h(t)\|_{D,\gamma_0 }^{2}\Big) +\nu_0 \|\partial
_{z}h(t)\|_{D,\gamma_0 }^{2}\\
&\leq&
KC(\gamma_0 +\delta _{0})\left(\|\partial _{t}h(t)\|_{D,\gamma_0
}^{2}+\|h(t)\|_{D,\gamma_0 }^{2}+\|\partial
 h(t)\|_{D,\gamma_0- \frac{1}{2}}^{2}\right)\\
 &+& KC\|h(t)\|_{D,\gamma_0- \frac{1}{2}}^{2}
+ C\Big(\|h(t)\|_{D,\gamma_0- \frac{1}{2}}^{2}\\
&+&\|\partial_t  h(t)\|_{D,\gamma_0- \frac{1}{2}}^{2}+\gamma_0\|\partial_z  h(t)\|_{D,\gamma_0- \frac{1}{2}}^{2}+\delta_0\|\partial h(t)\|^2_{D,\gamma_0}\Big)\ .
\end{eqnarray*}
For $\gamma_0$ small enough, for $\delta_0$ small enough and $K$ sufficiently large we get

\begin{eqnarray*}
\frac12\frac{d}{dt}&\Big(&K(\|h(t)\|_{M,\gamma_0 }^{2}+\|\partial _{t}h(t)\|_{M,\gamma_0 }^{2})+\|\partial _{z}h(t)\|_{M,\gamma_0 }^{2}\Big)\\
&+&K\nu_0\Big( \|h(t)\|_{D,\gamma_0 }^{2}+\partial
_{t}h(t)\|_{D,\gamma_0 }^{2}\Big) +\nu_0 \|\partial
_{z}h(t)\|_{D,\gamma_0 }^{2}
\\
&\leq&
KC(\|\partial
 h(t)\|_{D}^{2}+\|h(t)\|_{D}^{2}),
\end{eqnarray*}
Then, as before, by using (\ref{ult?}) we can conclude that
\begin{eqnarray}
 \nu_0\int_0^\infty dt &\Big(&K\nu_0 \left(\|h(t)\|_{D,\gamma_0 }^{2}+\partial
_{t}h(t)\|_{D,\gamma_0 }^{2}\right) +\nu_0 \|\partial
_{z}h(t)\|_{D,\gamma_0 }^{2}\Big)\nonumber\\&\le& C(\|h(0)\|_{M,\gamma_0}^2+\|\partial h(0)\|_{M,\gamma_0}^2).
\label{ult1}
\end{eqnarray}
Finally,
 we let $\gamma =\gamma _{0}$ sufficiently small in
Lemma \ref{0weight}, while let $\gamma =\frac{1}{2}+\gamma _{0}$ in both
Lemmas \ref{tweight} and \ref{zweight}, while multiplying the first two by $K$. We get

\begin{eqnarray*}
&&\frac12\frac{d}{dt}\Big(K\left(\|h(t)\|_{M,\gamma_0 }^{2}+\|\partial _{t}h(t)\|_{M,\gamma_0+ \frac{1}{2} }^{2}\right)+\|\partial _{z}h(t)\|_{M,\gamma_0+ \frac{1}{2} }^{2}\Big)\\
&&\quad +K\left(\nu_0 \|h(t)\|_{D,\gamma_0 }^{2}+\nu_0\|\partial
_{t}h(t)\|_{D,\gamma_0+ \frac{1}{2} }^{2}\right)+\nu_0 \|\partial
_{z}h(t)\|_{D,\gamma_0+ \frac{1}{2} }^{2}\\
&&\quad \leq
KC(\gamma_0 + \frac{1}{2}+\delta _{0})\left(\|\partial h(t)\|_{D,\gamma_0
}^{2}+\|h(t)\|_{D,\gamma_0 }^{2}\right)\\
&& +C\left(\frac{1}{2}\|\partial_{z} h(t)\|^2_{D,\gamma_0}+  \|h(t)\|_{D,\gamma_0
}^{2}+\|\partial_t h(t)\|_{D,\gamma_0}^{2}+\delta_0\|\partial
h(t)\|_{D,\gamma_0+ \frac{1}{2}}^{2} \right),
\end{eqnarray*}

Then,
there is a large constant $K
$ such that, for $\delta_0$ small,  
\begin{eqnarray*}
&&\frac{d}{dt}\Big(K(\|\partial _{t}h\|_{M,\frac{1}{2}+\gamma_{0}}^{2}+K\|h\|_{M,\gamma _{0}}^{2})+\|\partial _{z}h\|_{M,\frac{1}{2}+\gamma_{0}}\Big)\\
&+&{\nu_0 K}\Big(\|\partial _{t}h\|_{D,\frac{1}{2}+\gamma _{0}}^{2}+\|h\|_{D,\gamma _{0}}^{2}\Big)+\nu_0\|\partial _{z}h\|_{D,\gamma_{0}+\frac{1}{2}}\\
&\leq& KC\left(\|\partial
 h\|_{D,\gamma _{0}}^{2}+\|h\|_{D,\gamma _{0}}^{2}\right) \ .
\end{eqnarray*}
Using (\ref{ult1}) and a standard continuity argument, we obtain:
\begin{eqnarray}
\sup_{0\leq t\leq \infty }&\big(&\|h(t)\|_{M,\gamma _{0}}+\|\partial  h(t)\|_{M,
\frac{1}{2}+\gamma _{0}}\big)\nonumber\\&\leq& C\big(\|h(0)\|_{M,\gamma _{0}}+\|\partial
 h(0)\|_{M,\frac{1}{2}+\gamma _{0}}\big)\   \label{gamma0}
\end{eqnarray}
and the apriori assumption (\ref{gammasmall}) is valid when $\|h(0)\|_{M,\gamma _{0}}+\|\partial
 h(0)\|_{M,\frac{1}{2}+\gamma _{0}}$ is sufficiently small.

We now turn back to (\ref{largek}). We want to control $\|h\|_{M}+\|\partial h\|_M$ but up to now  we only have a uniform bound on  $\|h\|_D+\|\partial h\|_D$. What is missing is a bound on $\|a_h\|$. But from (\ref{gamma0}) and an interpolation,
\begin{eqnarray*}
\|a_{h}^{(1)}\| &&\leq \|\zhal\partial _{z}a_{h}^{(1)}\|\\ &&\leq
C\|(1+z^{2})^{\frac{1}{2}+\gamma _{0}}\partial _{z}a_{h}^{(1)}\|^{\frac{1}{
1+2\gamma _{0}}}\times \|\partial _{z}a_{h}^{(1)}\|^{\frac{2\gamma _{0}}{
1+2\gamma _{0}}} \\
&&\leq C\{\|h(0)\|_{\gamma _{0}}+\|\partial  h(0)\|_{\frac{1}{2}+\gamma
_{0}}\}^{\frac{1}{1+2\gamma _{0}}}\|h\|_{D}^{\frac{2\gamma _{0}}{1+2\gamma
_{0}}}.
\end{eqnarray*}
As for $a_h^{(2)}$, and $\partial  a_h$, by Lemma \ref{hydro},  we conclude that they satisfy the
same inequality above with $\gamma_0=0$.
Therefore, let $E_{\gamma _{0}}=\{\|h(0)\|_{\gamma _{0}}^{2}+\|\partial
 h(0)\|_{\frac{1}{2}+\gamma _{0}}^{2}\}$
\begin{equation*}
\{\|h\|_{D}^{2}+\|\partial  h\|_{D}^{2}\}\geq CE_{0}^{-\frac{1}{2\gamma
_{0}}}\{\|h\|^2+\|\partial  h\|^2\}^{\frac{1+2\gamma _{0}}{2\gamma _{0}}}.
\end{equation*}
We thus conclude that:
\begin{eqnarray*}
&&\frac{d}{dt}\{\|\partial _{z}h\|^{2}_M+K(\|h\|^{2}_M+\|\partial _{t}h\|^{2}_M)\}\\&&+CE_{0}^{-\frac{1}{2\gamma
_{0}}}\{\|\partial _{z}h\|^{2}_M+K(\|h\|^{2}_M+\|\partial _{t}h\|^{2}_M)\}^{1+\frac{1}{2\gamma _{0}}}\leq 0.
\end{eqnarray*}
Denoting $y(t)\equiv \|\partial _{z}h\|^{2}_M+K(\|h\|^{2}_M+\|\partial _{t}h\|^{2}_M)$, we have
\begin{equation*}
y^{\prime }y^{-1-\frac{1}{2\gamma _{0}}}\leq -CE_{0}^{-\frac{1}{2\gamma _{0}}
}.
\end{equation*}
Integrating over $0$ and $t$, we deduce
\begin{equation*}
\frac{1}{2\gamma _{0}}\{y(0)\}^{-\frac{1}{2\gamma _{0}}}-\frac{1}{2\gamma
_{0}}\{y(t)\}^{-\frac{1}{2\gamma _{0}}}\leq -CE_{0}^{-\frac{1}{2\gamma _{0}}
}t.
\end{equation*}
Hence from $y(0)\leq E_{0}$ we obtain
\begin{eqnarray*}
\frac{1}{2\gamma _{0}}\{y(t)\}^{-\frac{1}{2\gamma _{0}}} &\geq &t\frac{C}{
2\gamma _{0}}E_{0}^{-\frac{1}{2\gamma _{0}}}+\{y(0)\}^{-\frac{1}{2\gamma _{0}
}} \\
&\geq &\{t\frac{C}{2\gamma _{0}}+1\}E_{0}^{-\frac{1}{2\gamma _{0}}}\ 
\end{eqnarray*}
and the proof is completed by solving for $y(t)$. \qed

\bigskip

\noindent {\bf Acknowledgements}:{ The authors thank their institutions for support of the collaborations in this project. R.M. and R. E. are supported  in part by MIUR, INDAM-GNFM, and Y. G. is supported in part by NSF grant 0603615. R. E. and R. M. would like to thank the kind hospitality of Brown University, where this work started.}


\bigskip

\begin{thebibliography}{99}

\bibitem{As} \textsc{K. Asano}: \textit{private communication}.
\vskip.1cm
\bibitem{BL} \textsc{S. Bastea and J. L. Lebowitz}:\\ {Spinodal decomposition in binary gases.}, \textit{Phys.
Rev. Lett.} \textbf {78}, pp. 3499-3502 (1997).
\vskip.1cm
\bibitem {BELM} \textsc{S. Bastea, R. Esposito, J. L. Lebowitz and R.Marra}:\\ {
Binary fluids with
long range segregating interaction I:
derivation of kinetic and hydrodynamic equations}, \textit{ Jour. Stat. Phys.}, {\bf 101}, 1087--1136  (2000);\\ {Hydrodynamics of binary fluid phase segregation}, \textit{Phys. Rev. Letters}, {\bf 89} 235701-04 (2002);\\ {Sharp interface motion of a fluid binary mixture}, \textit{Jour. Stat. Phys.}, {\bf 124} pp. 445--483  (2006)
\vskip.1cm
\bibitem{CCELM1} \textsc{E. A. Carlen, C. C. Carvahlo, R. Esposito, J. L. Lebowitz, R. Marra}:\\
{Free energy minimizers for a two-species model with segregation and liquid-vapor transition},
\textit{Nonlinearity} {\bf 16}, 1075--1105 (2003).
\vskip.1cm
\bibitem{CCELM2} \textsc{E. A. Carlen, C. C. Carvahlo, R. Esposito, J. L. Lebowitz, R. Marra}:\\  { Displacement convexity and minimal fronts at phase boundaries},\textit{preprint} (2007).
\vskip.1cm
\bibitem{CCELM3} \textsc{E. A. Carlen, C. C. Carvahlo, R. Esposito, J. L. Lebowitz, R. Marra}:\\  \textit{unpublished}.
\vskip.1cm
\bibitem{CCO}
\textsc{E. A. Carlen, M.C. Carvalho, and E. Orlandi}:\\
{Algebraic rate of decay for the excess free
energy and stability of fronts for a non-local phase
kinetics equation with a conservation law, I}, \textit{Jour. Stat. Phys.}, {\bf 95}, pp 1069-1117 (1999);\\ {Algebraic rate of decay for the excess free energy and stability of fronts for a
non-local phase kinetics equation with a conservation law, II},
\textit{Comm. Par. Diff. Eq.}  {\bf 25}, pp 847-886  (2000).
\vskip.1cm
\bibitem{Guo} \textsc{Y. Guo}:\\ {The Boltzmann equation in the whole space}, {\textit Indiana Univ. Math. Jour.}, \textbf{53} pp. 1081Ð-1094 (2004).
\vskip.1cm
\bibitem{Guo1} \textsc{Y. Guo}:\\ {The Vlasov-Poisson-Boltzmann system near vacuum},  \textit{Comm. Math. Phys.} \textbf{218}, pp. 293Ð-313 (2001);\\ {The Vlasov-Maxwell-Boltzmann system near
Maxwellians},  \textit{Invent. math.}, \textbf{153}, pp. 593Ð-630 (2003);\\  {The Vlasov-Poisson-Boltzmann system near Maxwellians}, \textit{Comm. Pure Appl. Math.} \textbf{55}, no. 9, pp. 1104Ð-1135   (2002). 
\vskip.1cm
\bibitem{SG} \textsc{R. M. Strain and Y. Guo}:\\  {Almost exponential decay near Maxwellians}, \textit{Comm. Par. Diff. Eq.} , \textbf{31}, pp 417-429, (2006).
\vskip.1cm
\bibitem{LB}\textsc{D. Liberzon and R.W. Brockett}:\\  {Spectral analysis of Fokker-Planck and related operators arising from linear stochastic differential equations}, \textit{Siam  J. Control Optim.},{\bf 38}, No. 5, 1453Ð-1467 (2000).
\vskip.1cm
\bibitem {LY} \textsc{T.-P.  Liu, T. Yang  and S.-H. Yu}:\\   {Energy method for the Boltzmann equation} \textit{Physica D}, \textbf{188} pp 178--192 (2004).
\vskip.1cm
\bibitem {Ma} \textsc{N. B. Maslova}:\\ Nonlinear Evolution Equations, Kinetic Approach, \textit{Series on Advances in Mathematics for Applied Sciences}  Vol. 10, World Scientific (1993).
\vskip.1cm
\bibitem {MM}\textsc{G. Manzi and R. Marra}:\\ {Phase segregation and interface dynamics in kinetic systems,}
\textit{Nonlinearity}, \textbf{19}, pp 115-147  (2006);  {A kinetic model of interface motion.} \textit{International Journal of Modern Physics B},{\bf 18}, pp 1--10 (2004);\\ {Kinetic Modelling of Late Stages of Phase Separation} \textit{in ``Transport Phenomena and Kinetic Theory. Applications to Gases, Semiconductors, Photons, and Biological System''}, C. Cercignani and E. Gabetta Ed.s,
Birkhauser (2006).
\vskip.1cm
\bibitem{Uk}\textsc{S. Ukai}:\\ {On the existence of global solutions of a mixed problem for  the nonlinear Boltzmann equation}, \textit{Proc. Japan Acad.}, {\bf 50}, pp 179--184 (1974) 
\vskip.1cm
\bibitem {UYZ} \textsc{S. Ukai, T. Yang and H. Zhao}:\\ {Global solutions to the Boltzmann equation with external forces}, \textit{Analysis and Applications}, {\bf 3}, No. 2, pp. 157Ð-193  (2005).
\vskip.1cm
\bibitem{Vi} \textsc{C. Villani}:\\ {Hypocoercive diffusion operators }  \textit{in ``Proceedings of the  2006 International Congress of Mathematicians'' in Madrid.} Preprint (2006).
\vskip.1cm
\bibitem{Yu} \textsc{S.-H. Yu}:\\ {Hydrodynamic limits with shock waves of the Boltzmann equation}, \textit{Commun. Pure Appl. Math.}, {\bf 58}, pp 409--443 (2005).

\end{thebibliography}
\end{document}